\documentclass[review]{elsarticle}
\usepackage{ifpdf}
\usepackage{hyperref}
\usepackage{lineno}
\usepackage{subcaption}
\usepackage{amsmath}
\usepackage{amssymb}
\usepackage{siunitx}
\usepackage{graphicx}
\usepackage{float}
\usepackage{array, booktabs, adjustbox, makecell}
\usepackage{array}
\usepackage [english]{babel}
\usepackage [autostyle, english = american]{csquotes}

\MakeOuterQuote{"}

\modulolinenumbers[5]

\journal{Nuclear Instruments and Methods in Physics Research Section A}

\begin{document}

\begin{frontmatter}
\title{Application of deconvolution to recover frequency-domain multiplexed detector pulses}

\author[label1]{M. Mishra\corref{cor1}}
\address[label1]{Department of Nuclear Engineering, North Carolina State University, Raleigh, NC 27695}
\cortext[cor1]{Corresponding author}
\ead{mmishra4@ncsu.edu}

\author[label1]{J. Mattingly}
\author[label2]{R.M. Kolbas}
\address[label2]{Department of Electrical and Computer Engineering, North Carolina State University, Raleigh, NC 27695}




\begin{abstract}
Multiplexing of radiation detectors reduces the number of readout channels, which in turn reduces the number of digitizer input channels for data acquisition. We recently demonstrated frequency domain multiplexing (FDM) of pulse mode radiation detectors using a resonator that converts the detector signal into a damped sinusoid by convolution. The detectors were given unique "tags" by the oscillation frequency of each resonator. The charge collected and the time-of-arrival of the detector pulse were estimated from the corresponding resonator output in the frequency domain. 

In this paper, we demonstrate a new method to recover the detector pulse from the damped sinusoidal output by deconvolution. Deconvolution converts the frequency-encoded detector signal back to the original detector pulse. We have developed a new prototype FDM system to multiplex organic scintillators based on convolution and deconvolution. Using the new prototype, the charge collected under the anode pulse can be estimated from the recovered pulse with an uncertainty of about 4.4 keVee (keV electron equivalent). The time-of-arrival can be estimated from the recovered pulse with an uncertainty of about 102 ps. We also used a CeBr\textsubscript{3} inorganic scintillator to measure the Cs-137 gamma spectrum using the recovered pulses and found a standard deviation of 13.8 keV at 662 keV compared to a standard deviation of 13.5 keV when the original pulses were used. Coincidence measurements with Na-22 using the deconvolved pulses resulted in a timing uncertainty of 617 ps compared to an uncertainty of 603 ps using the original pulses. Pulse shape discrimination was also performed using Cf-252 source and EJ-309 organic scintillator pulses recovered by deconvolution. A figure of merit value of 1.08 was observed when the recovered pulses were used compared to 1.2 for the original pulses.

\end{abstract}


\end{frontmatter}

\linenumbers
\section{Introduction}
Multiplexing a large number of radiation detectors is useful when it is necessary to reduce the number of readout channels. FDM of pulse mode radiation detectors is based on the conversion of the detector signal into a damped sinusoid by a resonator via convolution. Each detector to be multiplexed is connected to a resonator with a unique oscillation frequency. Multiple resonators are connected to a single digitizer input channel by a fan-in circuit. A prototype readout system based on FDM by convolution was previously developed to multiplex organic scintillator detectors \cite{Mishra2018}. The frequency of the damped sinusoid identifies the detector number, and it was shown that the charge collected under the pulse could be estimated from the resonator signal amplitude. It was also shown that the time-of-arrival of the detector pulse could be estimated by applying constant fraction discrimination to the leading edge of the resonator signal. 

In this paper, we propose to convert the resonator output back to the original detector pulse that produced it by deconvolution. The pulse recovered through deconvolution can be used to estimate the charge collected and the time-of-arrival using standard charge integration and time-pickoff methods. Furthermore, for detectors that exhibit variations in pulse shape, particle identification can be implemented by applying pulse shape discrimination to the recovered pulse. This deconvolution method is intended for single-event multiplexing where the probability of more than one multiplexed detector producing signals in a single record is small. Our multiplexer system has been particularly designed to acquire data for the radiation detection experiments with low count rate ($\lesssim$ $10^{4}$ counts per second).

\subsection{FDM by convolution/deconvolution}
FDM of radiation detectors has been implemented in the past for transition-edge sensors (TES) using amplitude modulation of carrier current signal \cite{Irwin2005}. FDM is a technique by which the individual signals are transmitted in a single channel by shifting each signal into a separate frequency band. All the amplitude-modulated carrier currents flowing through their respective TESs are summed into a single channel and finally demodulated to recover each individual TES signal \cite{Smecher2012,Lanting2005,DenHartog2012,VanDerKuur2009}. This concept is analogous to the transmission of multiple analog voice signals over a single channel by shifting each signal to a unique frequency subband in a frequency division multiplexed telephone system \cite{Mitra2001}. 

\begin{figure}[H]
  \includegraphics[width=1.265\linewidth]{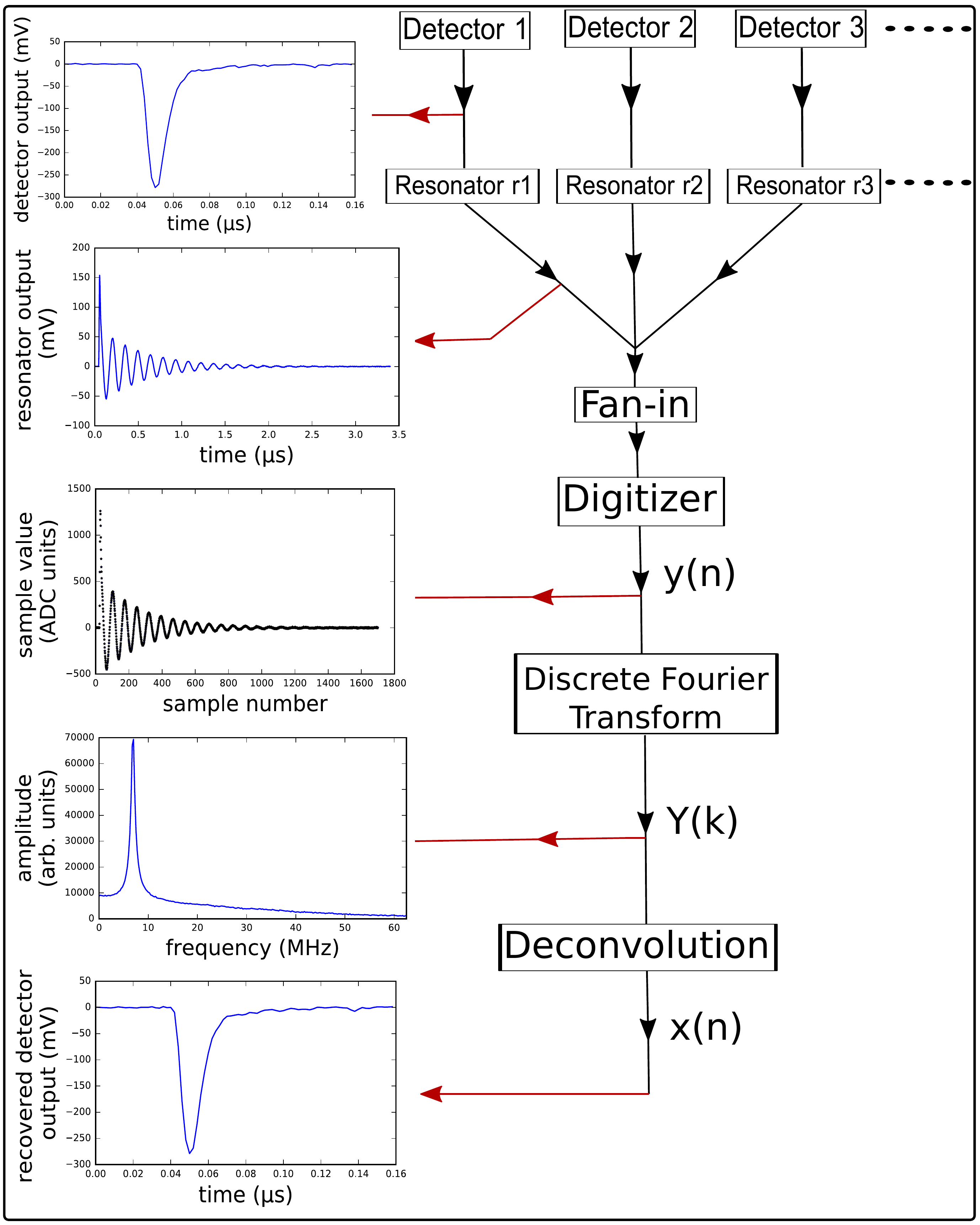}
  \caption{Block diagram of a frequency-domain multiplexer system using convolution/deconvoluiton.}
  \label{fig:fig1}
\end{figure}

Amplitude modulation is not practical for pulse mode detectors, so FDM has been implemented using convolution as illustrated by Fig. \ref{fig:fig1}. Each detector is connected to a resonator with a particular oscillation frequency. When one of the multiplexed detectors (e.g. detector 1 shown in Fig. \ref{fig:fig1}) emits a pulse, the corresponding damped sinusoid produced by its resonator r1 is passed into a single digitizer input channel by a fan-in circuit. The digitized output y(n) is the convolution between the impulse response $h_{r1}(n)$ of the resonator r1 in series with the fan-in circuit and the detector input x(n) \cite{dsp1}: 
\begin{equation}
  y(n) = \sum_{m=0}^{n} x(m)h_{r1}(n - m) \qquad n=0,1,...,N-1
\end{equation}
where N is the record length of the digitizer. The discrete Fourier transform of y(n) is given by
\begin{equation}
  Y(k) = \sum_{n=0}^{N-1} y(n)exp(\frac{-j2\pi kn}{N}) \qquad k=0,1,...,N-1
\end{equation}
In the frequency domain, the fan-in output $Y(k)$ is equivalent to the product of the impulse response $H_{r1}(k)$ and the detector pulse $X(k)$ given by
\begin{equation}
  Y(k) = H_{r1}(k)X(k) \qquad k=0,1,...,N-1
\end{equation}
The detector pulse $X(k)$ can be recovered from the damped sinusoid $Y(k)$ by deconvolution,
\begin{equation}
  X(k) = Y(k)/H_{r1}(k) 
\end{equation}
The discrete-time detector signal x(n) can finally be recovered by performing inverse Fourier transform on X(k): 
\begin{equation}
  x(n) = \frac{1}{N}\sum_{k=0}^{N-1} X(k)exp(\frac{j2\pi kn}{N}) 
\end{equation}

\section{Circuit design}
\begin{figure}[H]
  \centering
  \begin{subfigure}[t]{0.45\linewidth}
    \includegraphics[width=\linewidth]{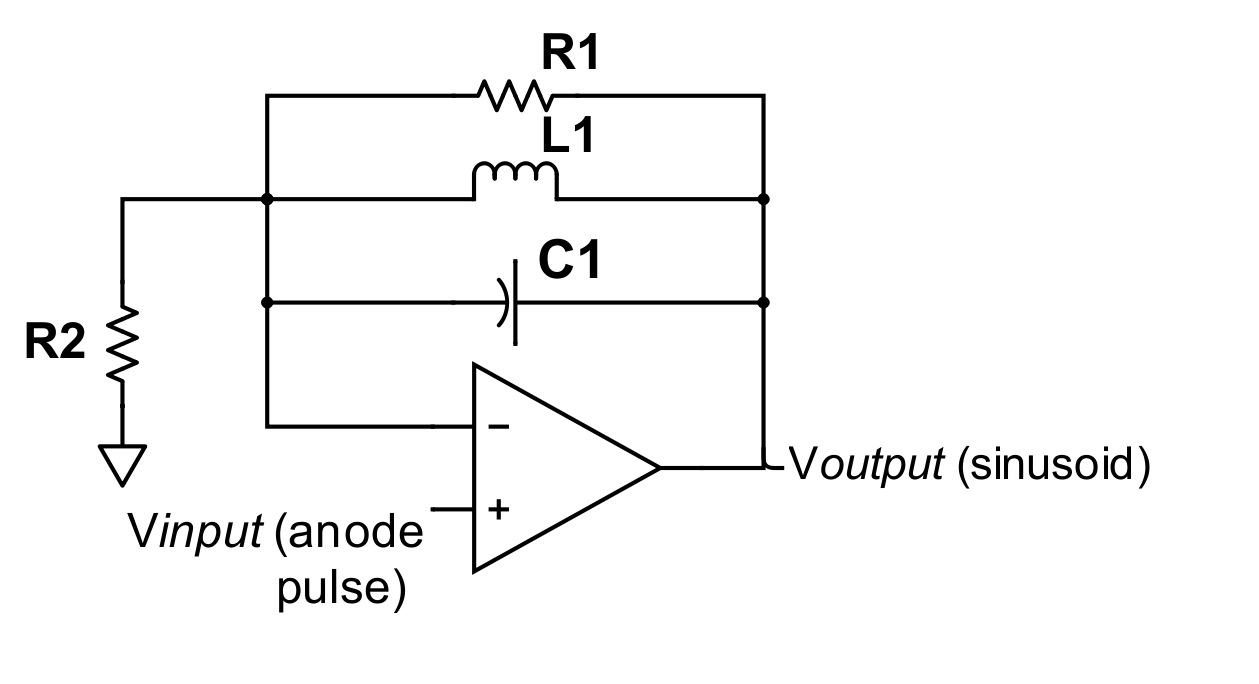}
     
    \caption{The resonator circuit, where $L_{1}$ and $C_{1}$ determine the resonant frequency and  $R_{1}$ determines the decay time of the damped sinusoid.}
    \label{fig:osc1}
  \end{subfigure}
  \hfill
  \begin{subfigure}[t]{0.45\linewidth}
    \includegraphics[width=\linewidth]{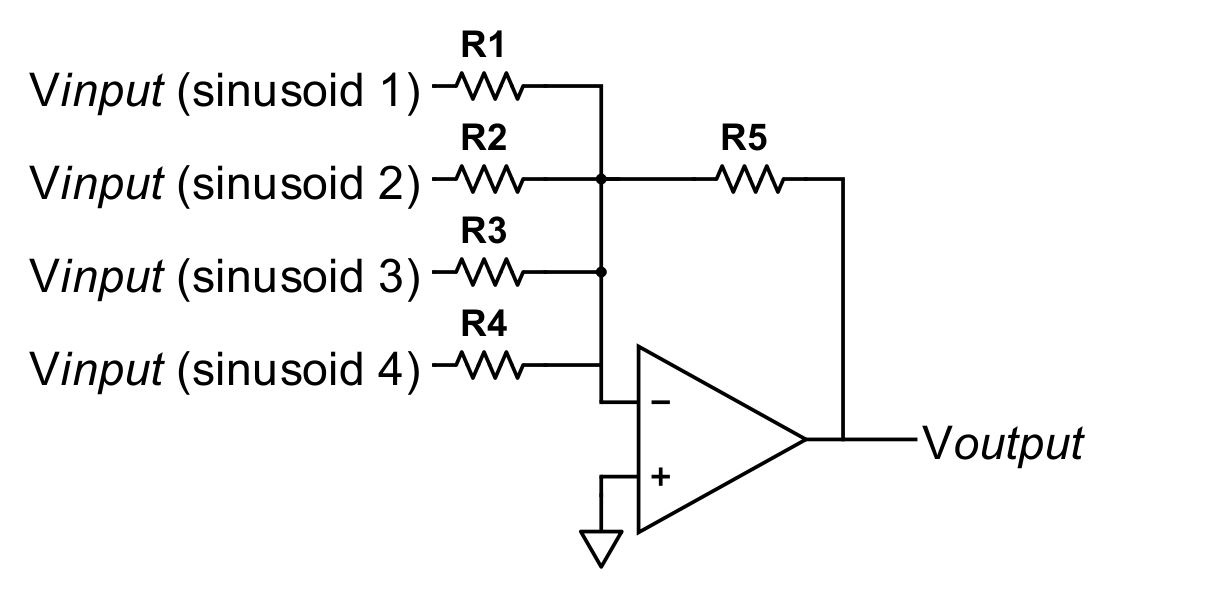}
    
    \caption{The fan-in circuit ($\frac{R_5}{R_1}=\frac{R_5}{R_2}=\frac{R_5}{R_3}=\frac{R_5}{R_4}=2$).}
    \label{fig:fan1}
  \end{subfigure}
  \caption{The electrical schematic of the circuits.}
  \label{fig:schematic}
\end{figure}

The basic design of the resonator and fan-in circuits was described in detail in Ref.\cite{Mishra2018}. The Electrical schematics of the circuits are shown in Fig. \ref{fig:schematic}. In this section we will discuss the changes made to the original design to improve the signal-to-noise ratio of the fan-in output at high frequencies, which was necessary for the accurate reconstruction of the detector pulses. 

The new resonator design uses OPA690 \cite{opa690}, a high bandwidth operational amplifier. The new fan-in circuit uses OPA659 \cite{opa659}, a high bandwidth, low noise operational amplifier. A double sided printed circuit board was used to make the circuits, with signal traces on the top layer and a full ground plane on the bottom. The short traces along with the bottom ground plane ensured lowest possible impedance in the current return paths to eliminate electromagnetic interference. 

Two prototype resonator circuits with resonant frequencies of 7.00 and 15.25 MHz respectively were fabricated. Following the original design, the Q-factor was kept between 10 and 15 to maintain the sinusoidal decay time of less than 2.5 $\mu$s and a bandwidth of less than 2 MHz for both the resonators. Each resonator has a pass-through that permits the original signal to be simultaneously digitized. The pass-through allows us to characterize the accuracy of the pulses reconstructed by deconvolution; it is not a necessary feature of the resonator. The pass-through is 50 ohm terminated.
  
A fan-in circuit with a gain of 2 was fabricated. This circuit has four input channels same as the original design. 

\section{FDM setup}
\begin{figure}[H]
  \includegraphics[width=0.9\linewidth]{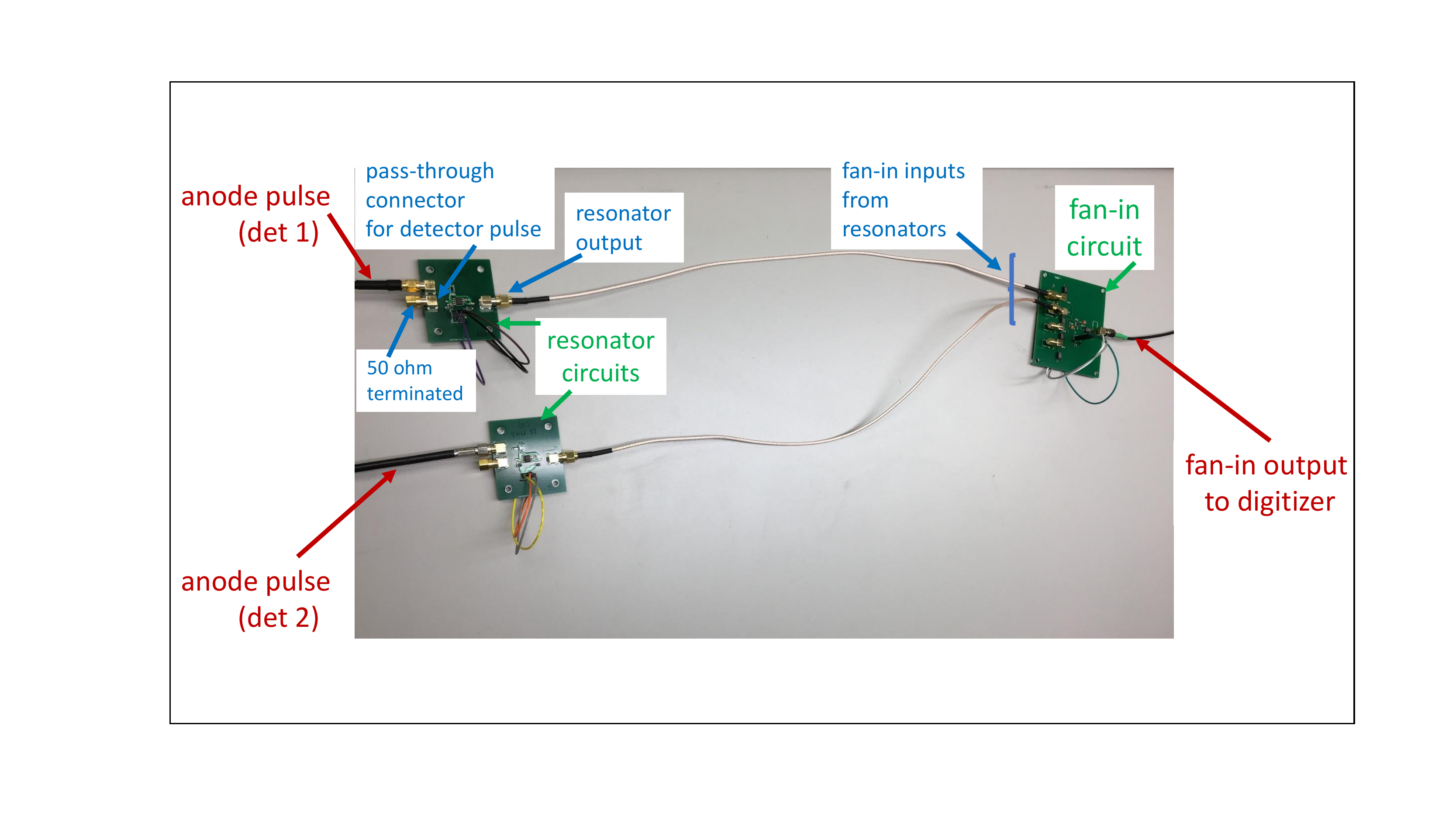}
  \caption{The FDM system. Two EJ-309 detectors were connected to two resonators, and the resonator outputs were finally combined by the fan-in circuit. For scale, the connectors shown are SMA connectors.}
  \label{fig:setup1}
\end{figure}

Two 7.6 cm x 7.6 cm EJ-309 detectors, each coupled to an Electron Tube 9821 KEB PMT, were connected to two resonators with resonant frequencies of 7.00 MHz and 15.25 MHz, shown in Fig. \ref{fig:setup1}. The detector signals acted as the inputs to the resonator circuits. The resulting damped sinusoids from the resonator outputs were combined using the fan-in circuit. The fan-in output was finally connected to a CAEN DT5730B, 14-bit, 500 MS/s digitizer. The digitizer saved a record length of 4 $\mu$s (2000 samples) for each channel when triggered. 

\section{Measurements and Results}
\subsection{Impulse response}
When one of the multiplexed detectors produces an event, the digitized damped sinusoidal output is deconvolved in the frequency domain using the impulse response of the corresponding resonator in series with the fan-in circuit. The impulse response of each resonator was measured by connecting it with the fan-in circuit separately. A 1 MHz, 1 Vpp noise signal from a function generator was used as input to the resonator. We simultaneously digitized the noise input via the resonator pass-through and the corresponding resonator output via the fan-in circuit. 

We used the cross-correlation technique to measure the impulse response of each resonator in series with the fan-in circuit. The cross-correlation between the output and the input sequence is given by \cite{dsp1}
\begin{equation}
  r_{yx}(m) = \sum_{l=0}^{m} h(l)r_{xx}(m - l) \qquad m=0,1,...,N-1
\end{equation}
where $r_{yx}$ is the cross-correlation of the fan-in output with the noise input, $r_{xx}$ is the autocorrelation of the noise input, and $h$ is the impulse response. The autocorrelation of an ideal white noise signal is a delta function and the corresponding cross-correlation output provides the impulse response. The corresponding relationship in the frequency domain is $S_{yx}(k) = H(k)S_{xx}(k)$, such that the impulse response can be estimated from
\begin{equation}
  H(k) = \frac{S_{yx}(k)}{S_{xx}(k)} \qquad k=0,1,...,N-1 
\end{equation}
where $H(k)$, $S_{yx}(k)$ and $S_{xx}(k)$ are the discrete Fourier tranforms of their respective counterparts in discrete-time domain. With N = 2,000, the range $0\leq k \leq \frac{N}{2}-1$ corresponds to a frequency range between 0 and 250 MHz for the sampling rate of 500 MS/s. 

\begin{figure}[H]
  \centering
  \begin{subfigure}[t]{0.7\linewidth}
    \includegraphics[width=\linewidth]{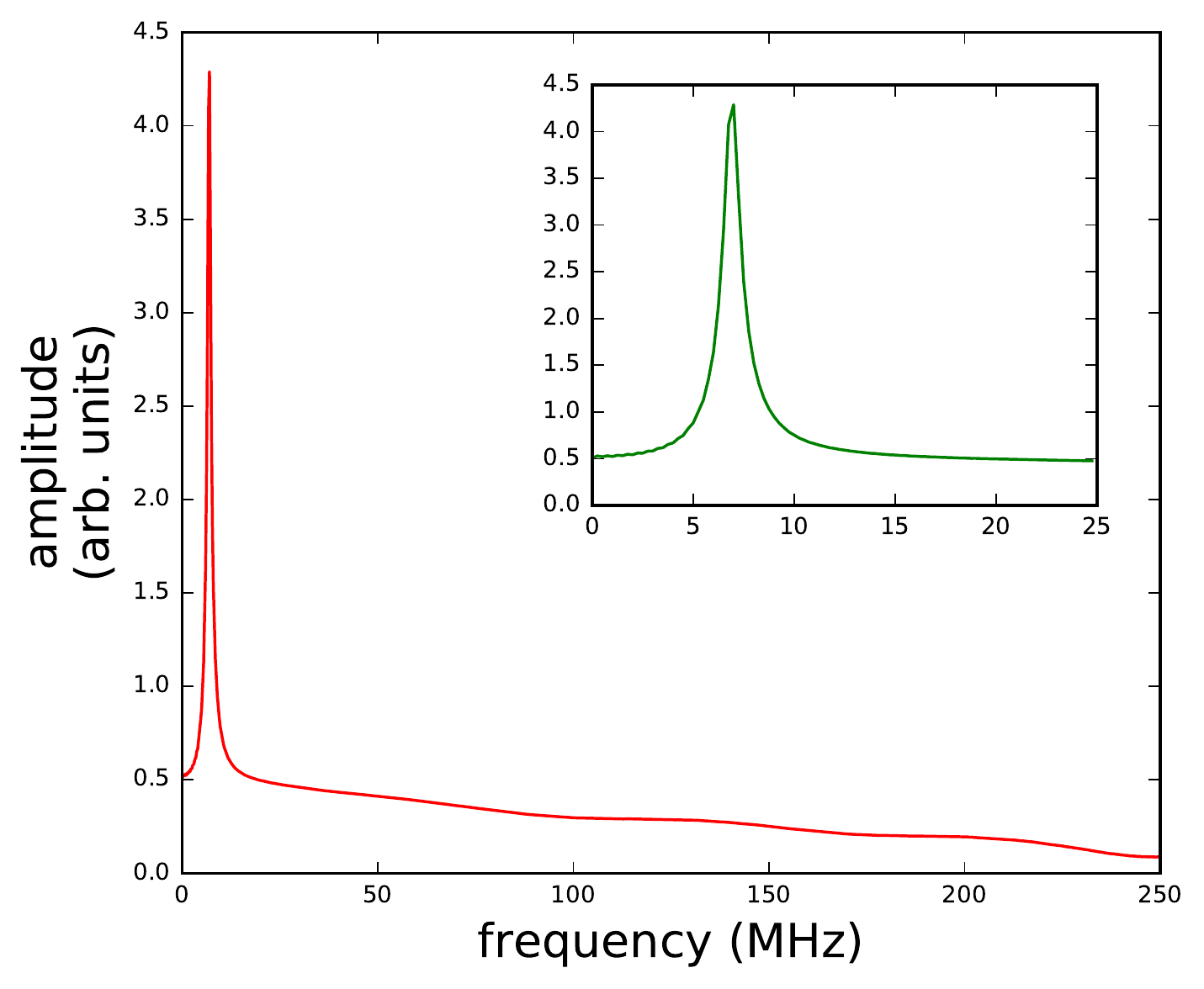}
     
    \caption{The impulse response of the 7.00 MHz resonator in series with the fan-in circuit in the frequency domain. The inset plot shows the peak at 7 MHz.}
    \label{fig:impres}
  \end{subfigure}
  \hfill
  \begin{subfigure}[t]{0.7\linewidth}
    \includegraphics[width=\linewidth]{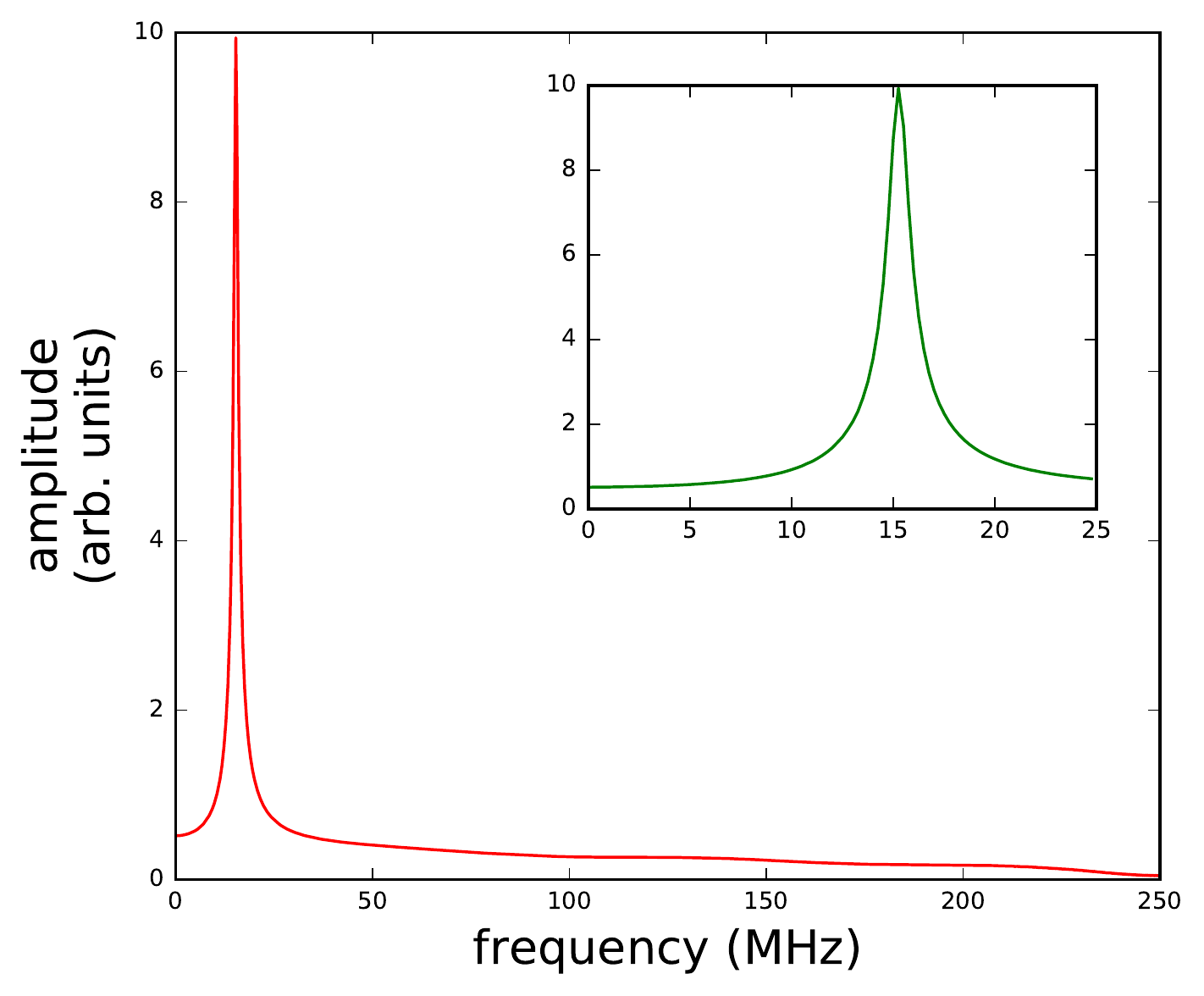}
    
    \caption{The impulse response of the 15.25 MHz resonator in series with the fan-in circuit in the frequency domain. The inset plot shows the peak at 15.25 MHz.}
    \label{fig:impres15}
  \end{subfigure}
  \caption{Resonator/fan-in impulse response.}
  \label{fig:impress1}
\end{figure}

We first acquired 10,000 records (each with 2,000 samples) of noise input and the corresponding fan-in output. We then computed the cross-correlation and the autocorrelation sequences for each of the 10,000 records. Subsequently, the average cross-correlation ($r_{yx}$) and the average autocorrelation ($r_{xx}$) were computed by averaging the individual cross-correlation and the autocorrelation sequences. The impulse response was finally computed by dividing the Fourier transform of the average cross-correlation ($S_{yx}$) by the Fourier transform of the average autocorrelation ($S_{xx}$). The impulse response of each resonator in series with the fan-in circuit is shown in Fig. \ref{fig:impress1} with resonant peaks at 7.00 MHz and 15.25 MHz respectively.

\subsection{Deconvolution}
Deconvolution can be performed by dividing the discrete Fourier transform of the output sequence Y(k) by the impulse response H(k) of the corresponding resonator in series with the fan-in circuit,
\begin{equation}
  X(k) = \frac{Y(k)}{H(k)} \qquad k=0,1,...,N-1  \qquad for \qquad N=2000
\end{equation}  
Fig. \ref{fig:impress1} shows that the amplitude of the impulse response above 200 MHz is close to zero because the fan-in output signal becomes comparable to the noise of the system. This results in the amplification of noise \cite{Smith} in the recovered signal X(k) at these high frequencies after deconvolution. This noise is filtered from the recovered signal X(k) by applying an optimized fourth order Butterworth low-pass filter with a cutoff frequency of 180 MHz.   

\subsection{Signal reconstruction}
\begin{figure}[H]
  \centering
  \begin{subfigure}[t]{0.66\linewidth}
    \includegraphics[width=\linewidth]{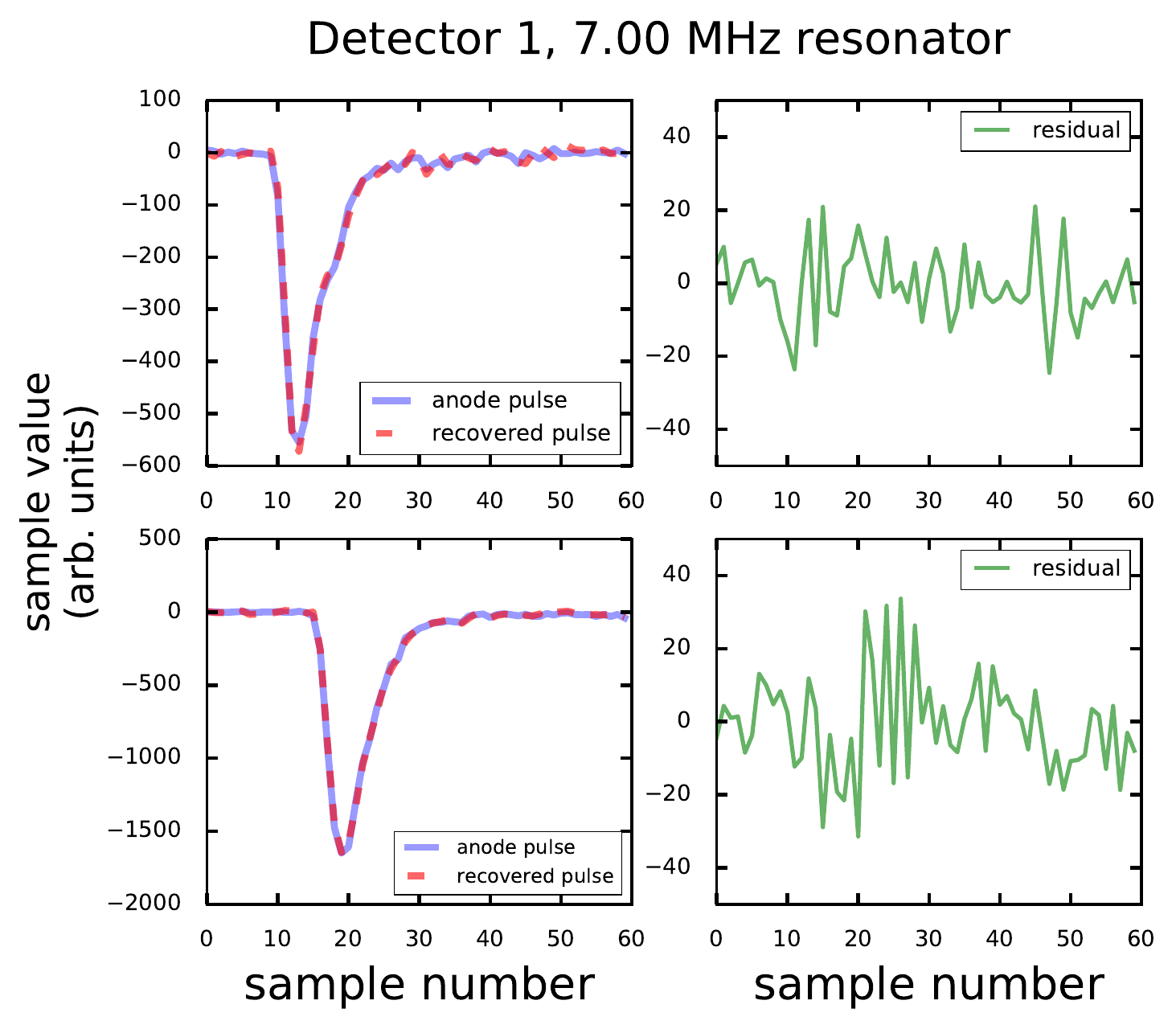}
     
    \caption{The comparison between the anode and the recovered pulses for the detector 1 connected to the 7.00 MHz resonator. The difference between the anode and the recovered pulses is shown on the right.}
    \label{fig:reconst1}
  \end{subfigure}
  \hfill
  \begin{subfigure}[t]{0.66\linewidth}
    \includegraphics[width=\linewidth]{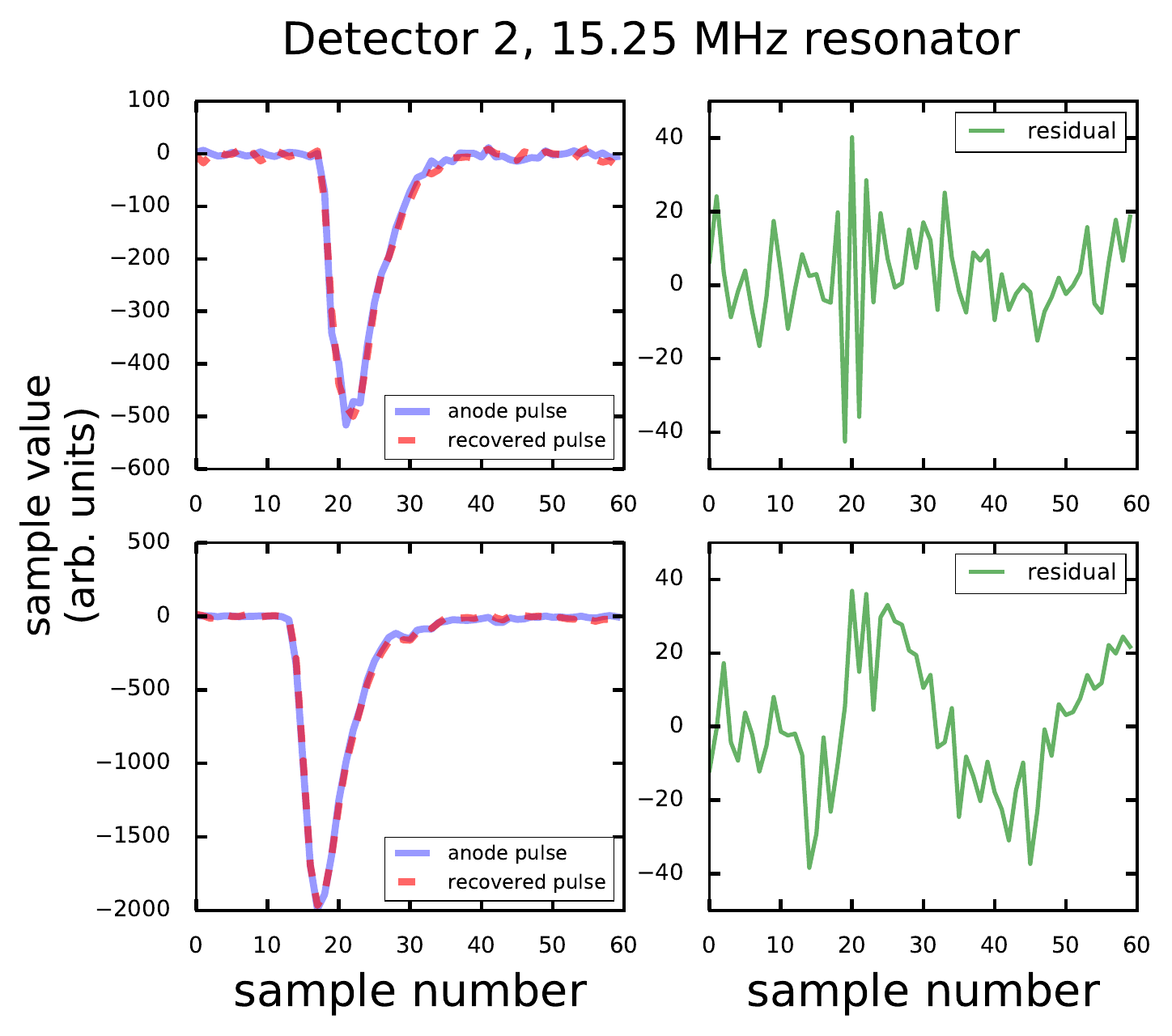}
    
    \caption{The comparison between the anode and the recovered pulses for the detector 2 connected to the 15.25 MHz resonator. The difference between the anode and the recovered pulses is shown on the right.}
    \label{fig:reconst2}
  \end{subfigure}
  \caption{Anode signal recovery.}
  \label{fig:reconst}
\end{figure}

Deconvolution was performed on the damped sinusoids when one of the multiplexed detectors fired in a particular record of digitized data. The anode pulses induced by a Cs-137 source were simultaneously digitized via the resonator pass-through and the damped sinusoids via the fan-in output. Fig. \ref{fig:reconst} compares the original anode pulses from each of the EJ-309s to the corresponding recovered pulses deconvolved from the damped sinusoids produced by each of the resonators. This comparison is shown for a low amplitude and a high amplitude pulse. 

\begin{figure}[H]
  \includegraphics[width=0.7\linewidth]{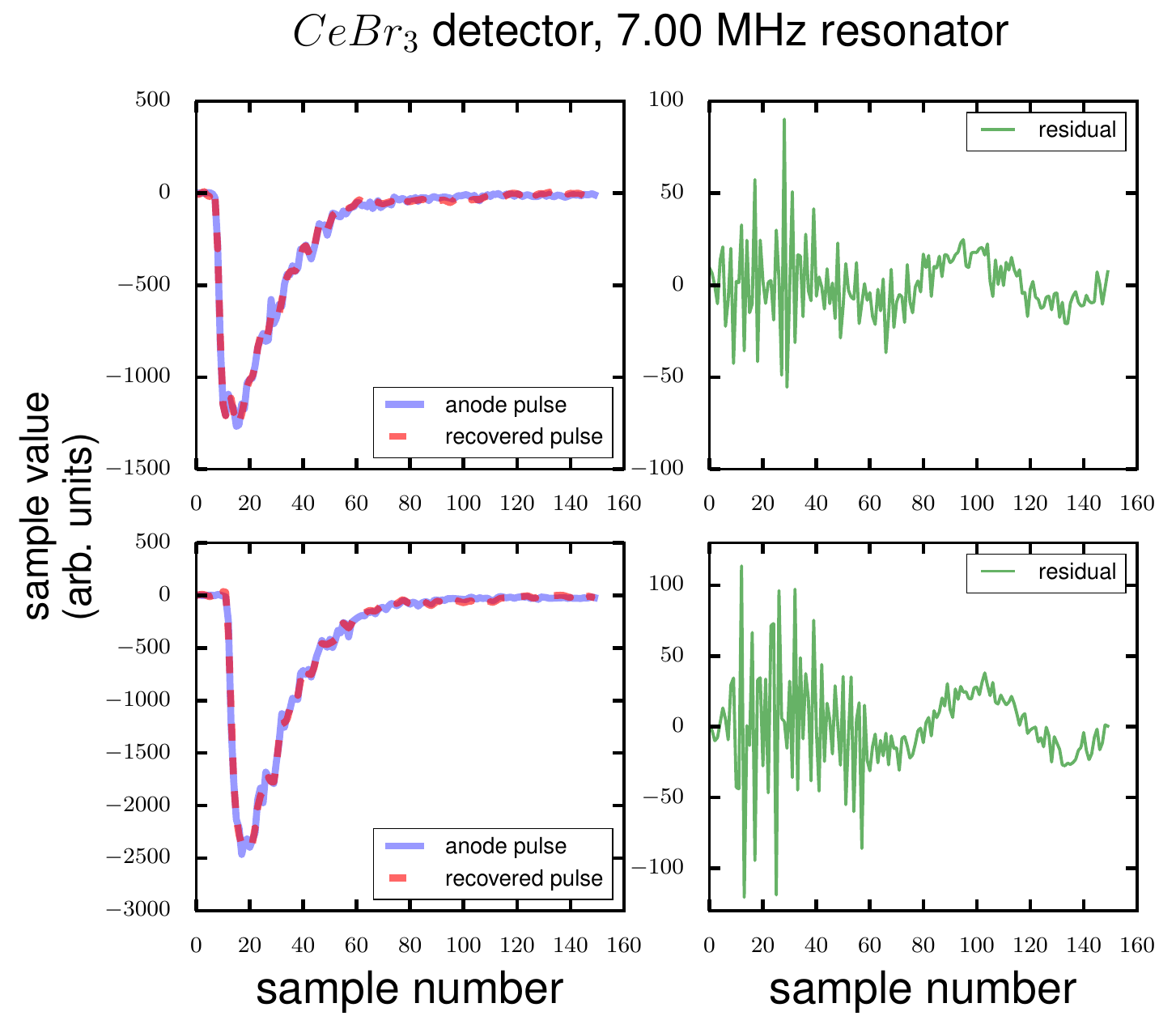}
  \caption{Anode pulse recovery using a CeBr\textsubscript{3} detector connected to the 7.00 MHz resonator in series with the fan-in circuit. The difference between the anode and the recovered pulses is shown on the right.}
  \label{fig:reconst11}
\end{figure}

We also used a CeBr\textsubscript{3} detector connected to the circuits. Fig. \ref{fig:reconst11} compares the anode pulses induced by a Cs-137 source to the corresponding recovered pulses.

\subsection{Charge estimation using the recovered pulse}
\begin{figure}[H]
  \centering
  \begin{subfigure}[!htb]{0.65\linewidth}
    \includegraphics[width=\linewidth]{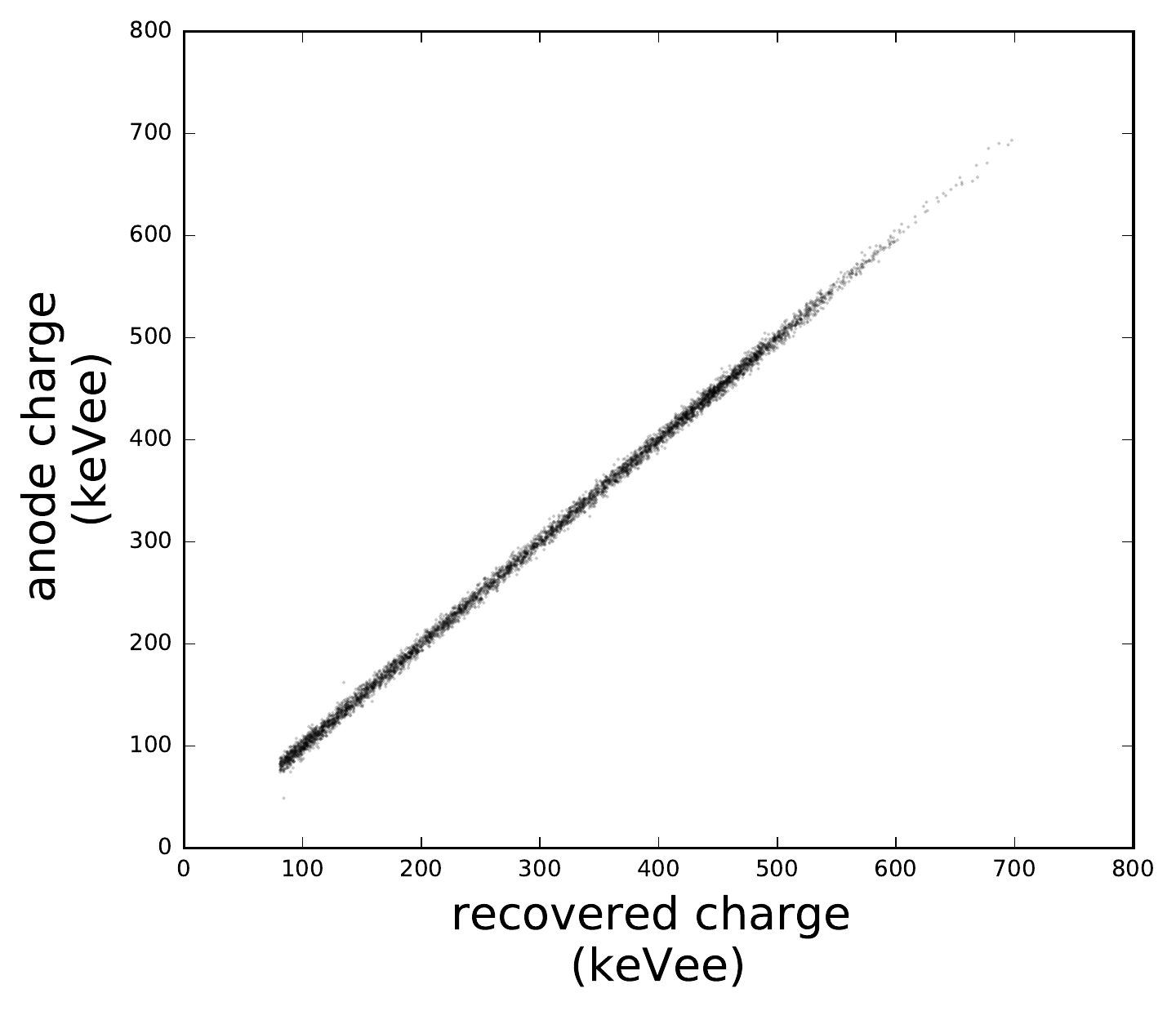}
     
    \caption{Scatter plot between the anode charge from EJ-309 plotted against the recovered charge using the 7.00 MHz resonator in series with the fan-in circuit.}
    \label{fig:ej_ch1}
  \end{subfigure}
  \hfill
  \begin{subfigure}[!htb]{0.65\linewidth}
    \includegraphics[width=\linewidth]{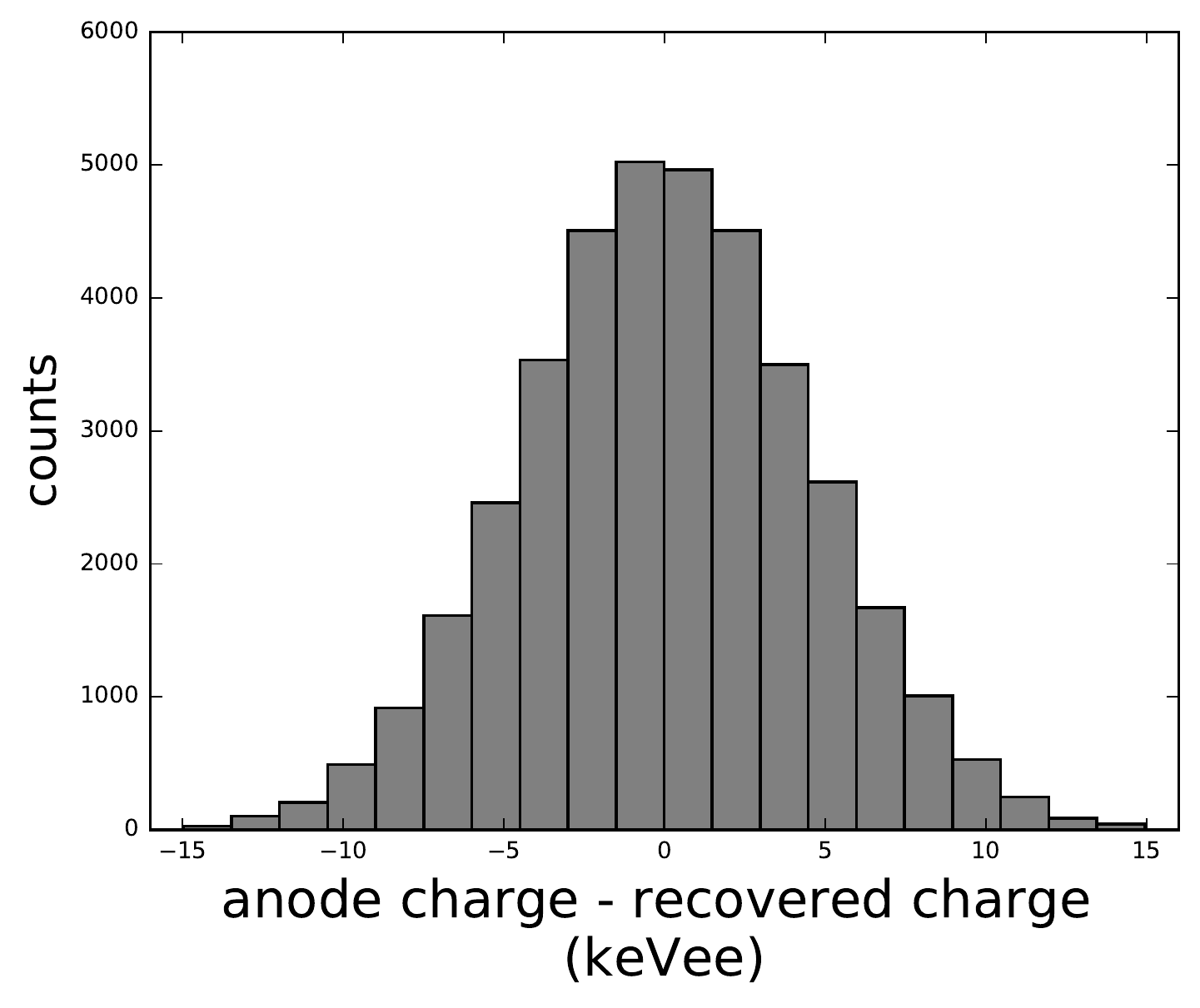}
    
    \caption{The events of the scatter plot in Fig. \ref{fig:ej_ch1} shown as the histogram of the difference between the anode charge and the recovered charge with a standard deviation of 4.4 $\pm$ 0.04 keVee.}
    \label{fig:ej_ch2}
  \end{subfigure}
  \caption{Estimation of the charge collected under the anode pulse from the recovered pulse using EJ-309.}
  \label{fig:ej_ch}
\end{figure}

The charge collected (i.e., the area under the anode pulse) from the EJ-309 detector can be estimated from the corresponding recovered pulse. Each resonator was connected to the fan-in circuit separately, and anode pulses were generated using a Cs-137 source. The charge collected under the anode pulse acquired via the resonator pass-through was compared to the charge collected under the recovered pulse deconvolved from the corresponding damped sinusoid. Fig. \ref{fig:ej_ch1} shows the scatter plot between the charge under the anode pulse and the charge under the recovered pulse. Fig. \ref{fig:ej_ch2} shows the events of the scatter plot as the histogram of the difference between the anode charge and the recovered charge. The standard deviation of 4.4 $\pm$ 0.04 keVee computed from the distribution is the uncertainty in the estimate of the anode charge from the recovered charge. Using the deconvolved pulses, the uncertainty in the estimate of the charge collected was reduced by more than half compared to the uncertainty from the amplitude of the damped sinusoid discussed in \cite{Mishra2018}.

\begin{figure}[H]
  \centering
  \begin{subfigure}[!htb]{0.65\linewidth}
    \includegraphics[width=\linewidth]{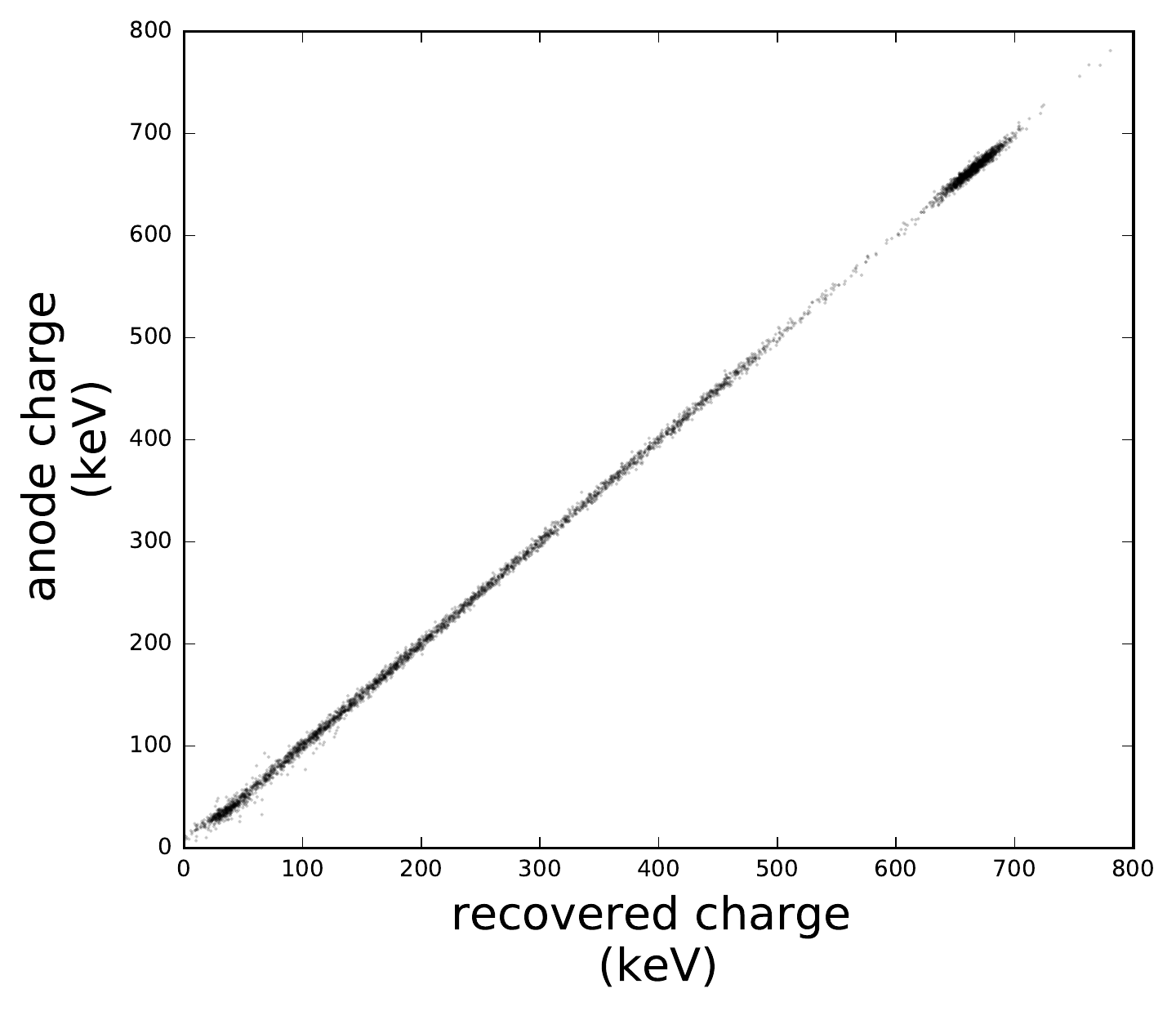}
     
    \caption{Scatter plot between the anode charge from CeBr\textsubscript{3} plotted against the recovered charge using the 7.00 MHz resonator in series with the fan-in circuit.}
    \label{fig:ce_ch11}
  \end{subfigure}
  \hfill
  \begin{subfigure}[!htb]{0.65\linewidth}
    \includegraphics[width=\linewidth]{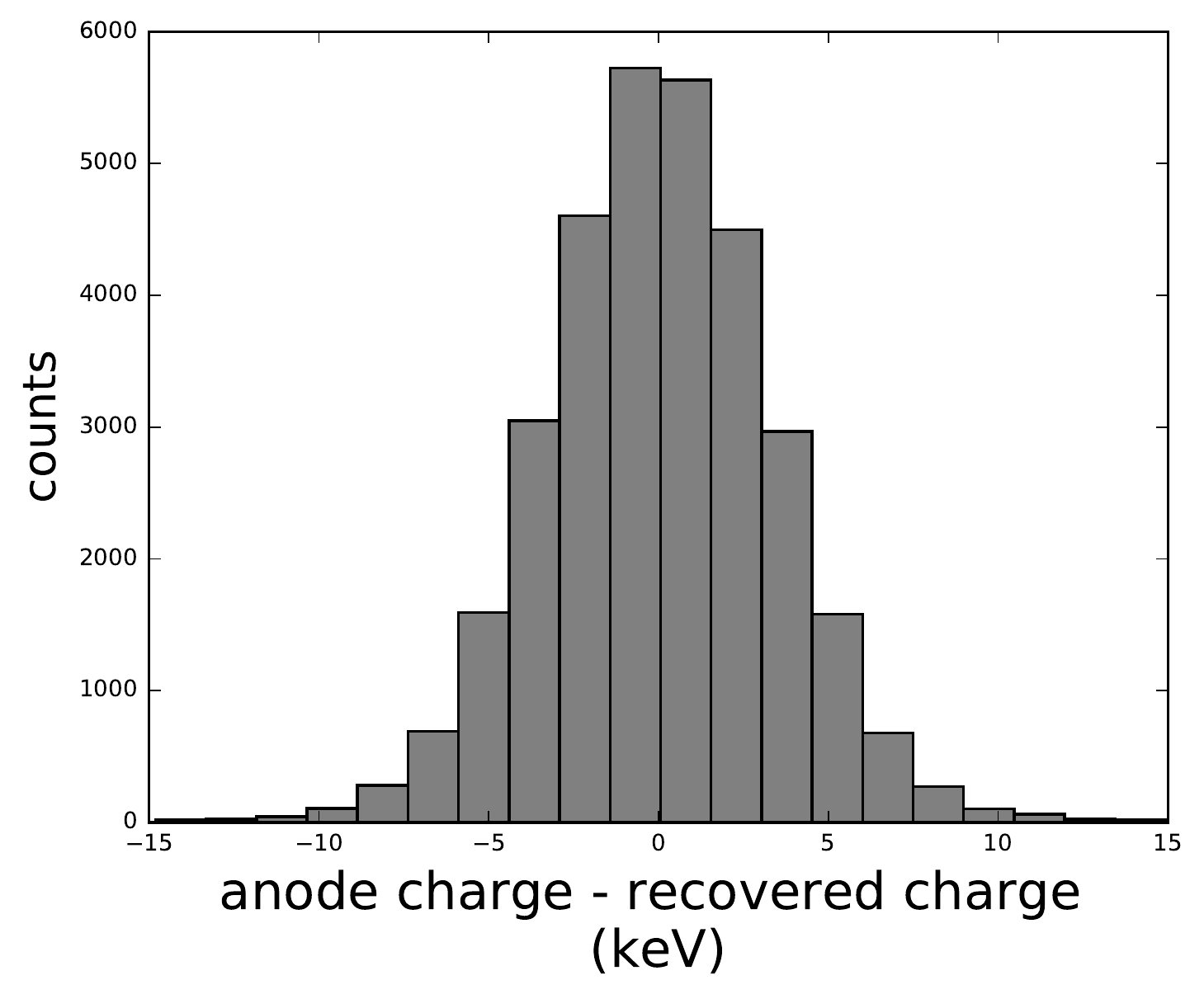}
    
    \caption{The events of the scatter plot in Fig. \ref{fig:ce_ch11} shown as the histogram of the difference between the anode charge and the recovered charge with a standard deviation of 3.2 $\pm$ 0.3 keV.}
    \label{fig:ce_ch22}
  \end{subfigure}
  \caption{Estimation of the charge collected under the anode pulse from the recovered pulse using CeBr\textsubscript{3}.}
  \label{fig:ce_ch1}
\end{figure}
\begin{figure}[H]
  \centering
  \begin{subfigure}[t]{0.55\linewidth}
    \includegraphics[width=\linewidth]{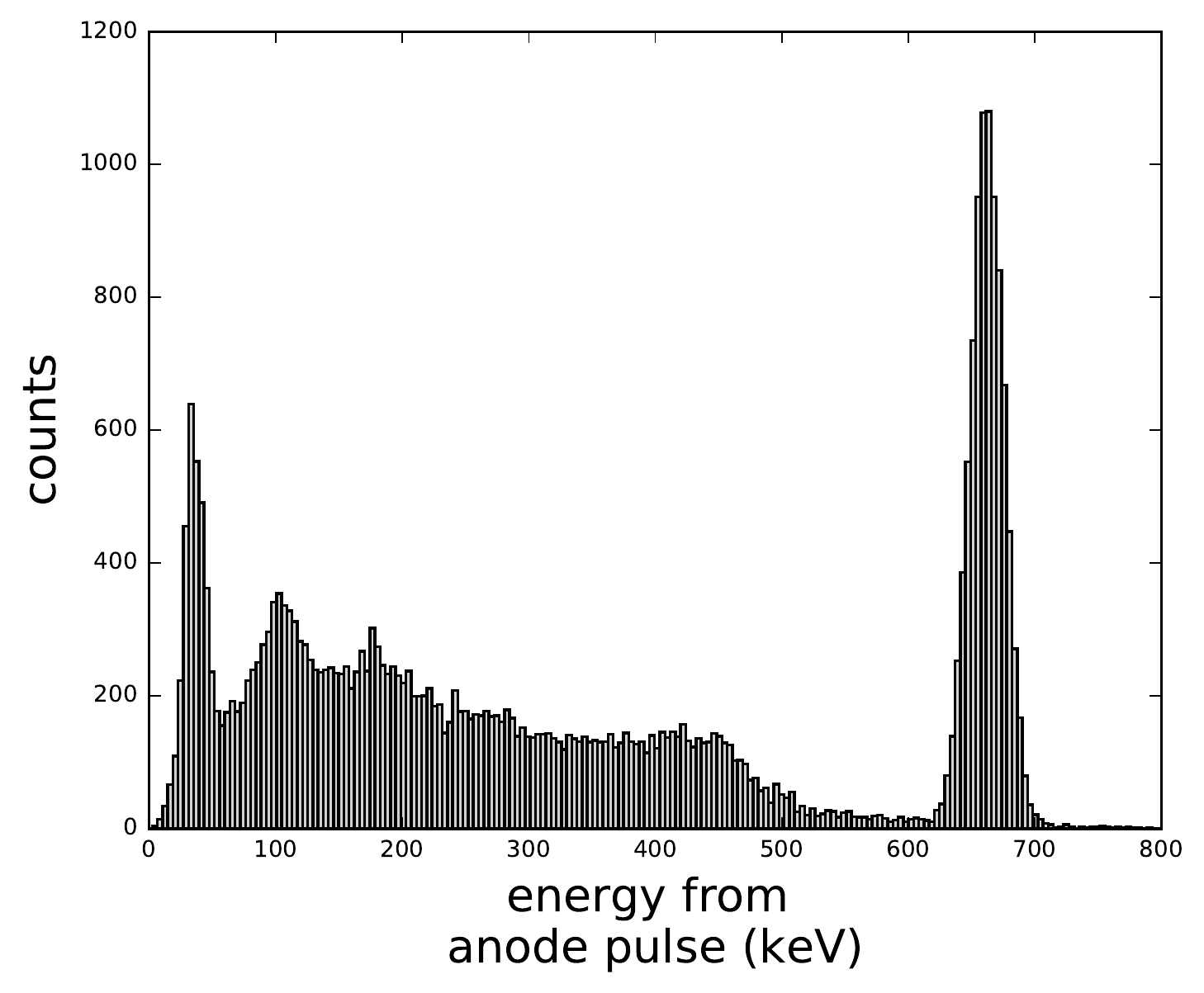}
     
    \caption{Pulse height spectrum measured from the anode pulses. The standard deviation at 662 keV is 13.5 $\pm$ 0.4 keV. }
    \label{fig:res211}
  \end{subfigure}
  \hfill
  \begin{subfigure}[t]{0.55\linewidth}
    \includegraphics[width=\linewidth]{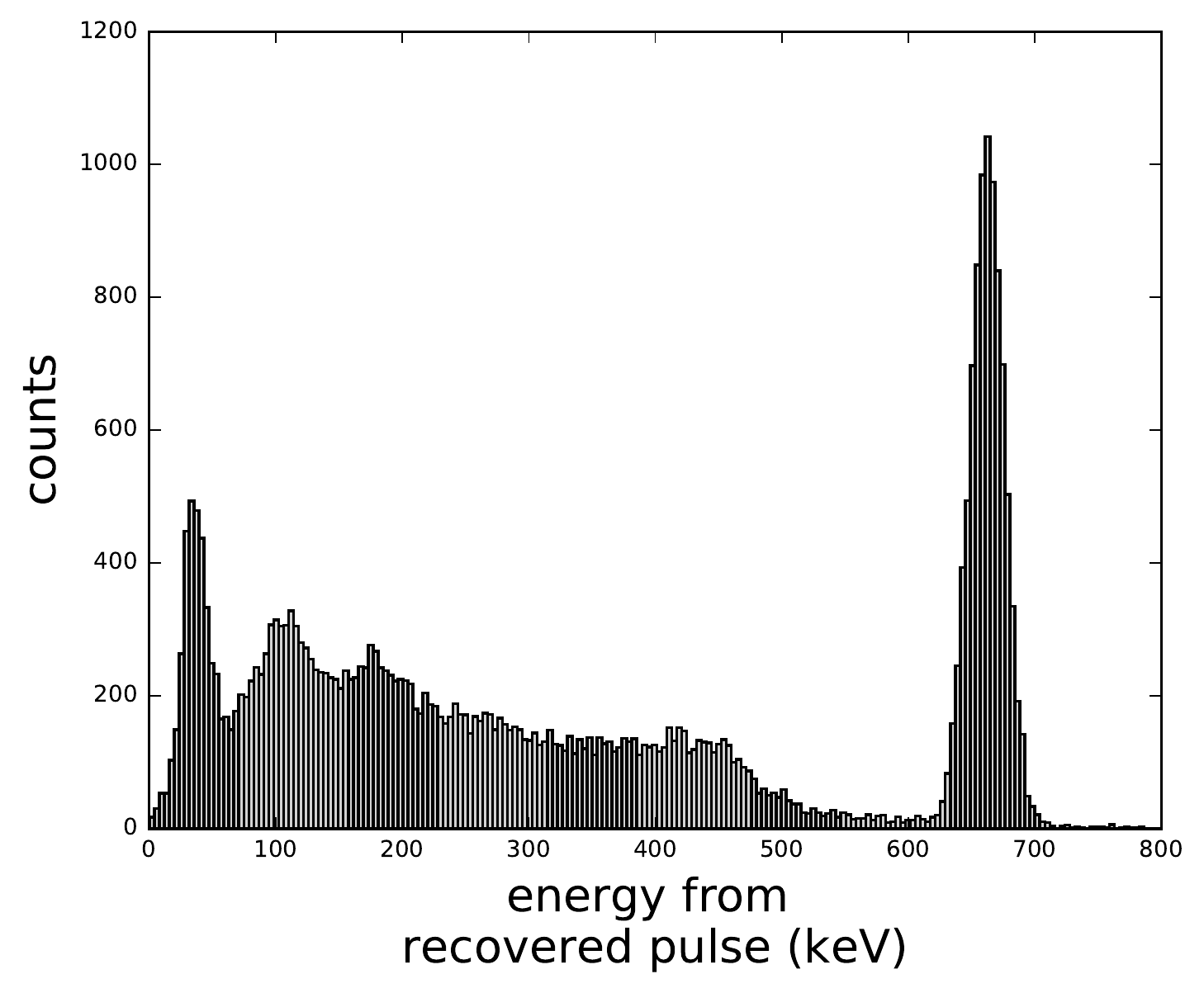}
    
    \caption{Pulse height spectrum measured from the recovered pulses. The standard deviation at 662 keV is 13.8 $\pm$ 0.4 keV.}
    \label{fig:res222}
  \end{subfigure}
  \caption{Comparison of $\gamma$-ray spectra from CeBr\textsubscript{3} measuring Cs-137 using the anode pulses and the recovered pulses respectively.}
  \label{fig:ress3}
\end{figure}

We also used a CeBr\textsubscript{3} detector connected to the 7.00 MHz resonator in series with the fan-in circuit to accumulate the pulse height spectrum using a Cs-137 source. The anode pulses were again digitized via the resonator pass-through and the corresponding damped sinusoids were digitized via the fan-in output. Fig. \ref{fig:ce_ch11} shows the scatter plot between the anode charge and the recovered charge, which is also shown as the histogram of the difference between the anode charge and the recovered charge in Fig. \ref{fig:ce_ch22}. The standard deviation of 3.2 $\pm$ 0.3 keV computed from the histogram is the uncertainty in the estimate of the anode charge from the recovered charge. Fig. \ref{fig:ress3} compares the pulse height spectrum using the anode pulses from the CeBr\textsubscript{3} detector to the spectrum using the deconvolved pulses from the corresponding damped sinusoids. The standard deviation of the photopeak at 662 keV for the spectrum using the anode pulses (Fig. \ref{fig:res211}) is 13.5 $\pm$ 0.4 keV while the standard deviation using the recovered pulses (Fig. \ref{fig:res222}) is 13.8 $\pm$ 0.4 keV. Note the area under the 662 keV photopeak is identical in both spectra. In both tests, the error in charge estimated from the recovered pulses had a mean of approximately zero, and the mean and standard deviation were independent of pulse amplitude. 

\subsection{Time pick-off using the recovered pulse}
\begin{figure}[H]
  \centering
  \begin{subfigure}[!htb]{0.65\linewidth}
    \includegraphics[width=\linewidth]{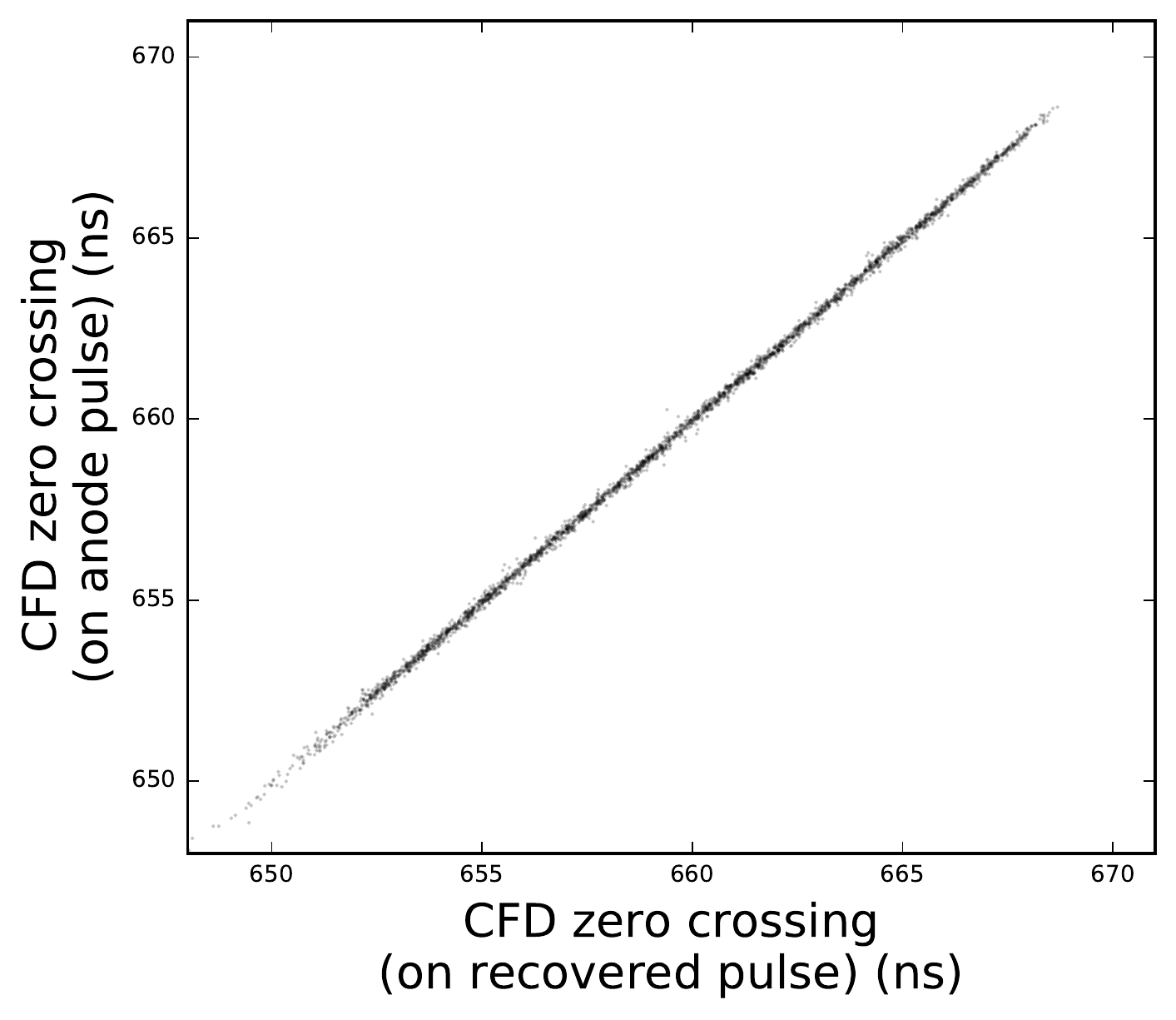}
     
    \caption{Scatter plot between time pick-off of the anode pulse from EJ-309 plotted against the time pick-off of the recovered pulse using the 7.00 MHz resonator in series with the fan-in circuit.}
    \label{fig:at_rt1}
  \end{subfigure}
  \hfill
  \begin{subfigure}[!htb]{0.65\linewidth}
    \includegraphics[width=\linewidth]{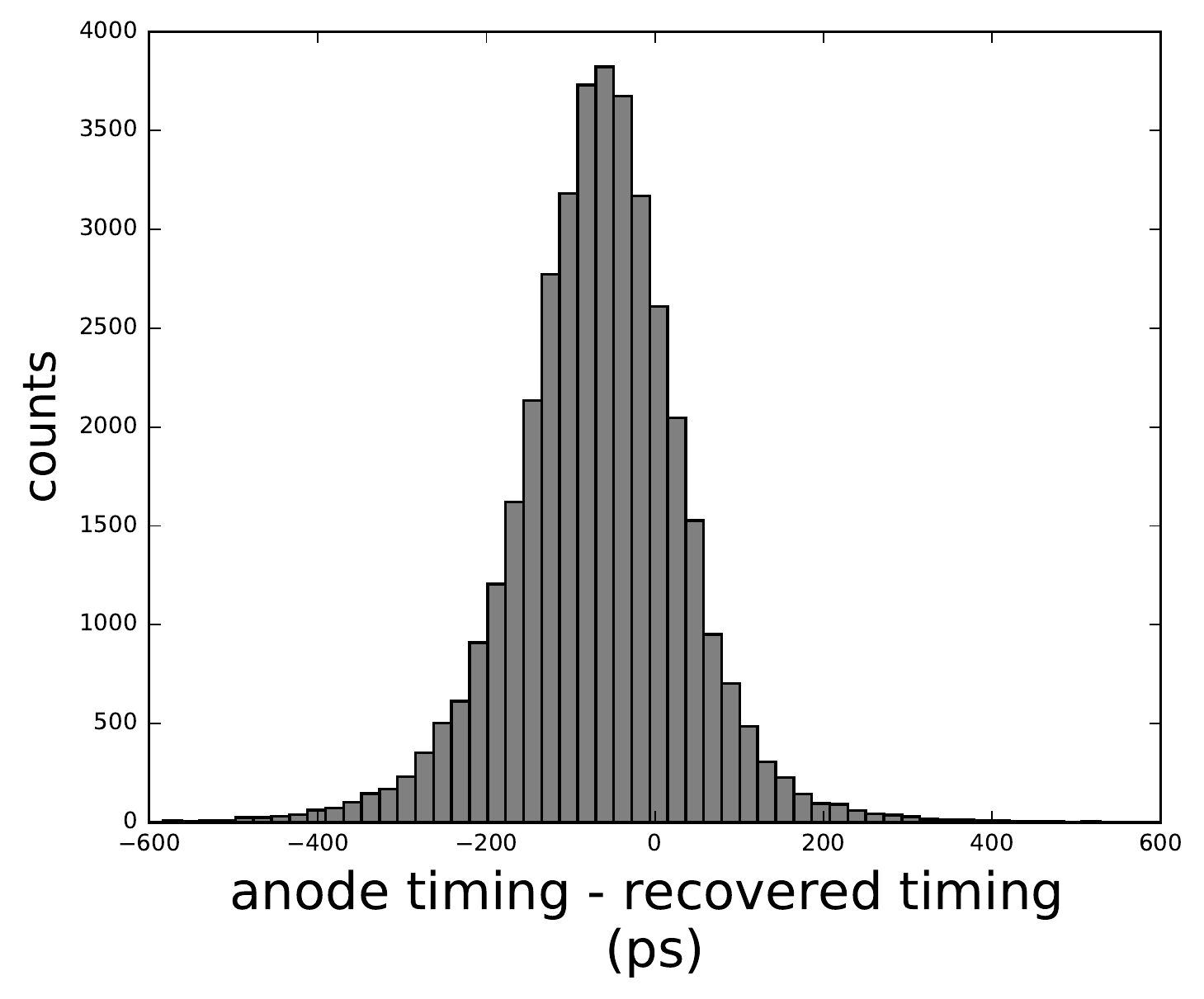}
    
    \caption{The events of the scatter plot in Fig. \ref{fig:at_rt1} shown as the histogram of the difference between the anode timing and the recovered timing with a standard deviation of 102 $\pm$ 1.5 ps.}
    \label{fig:at_rt2}
  \end{subfigure}
  \caption{Estimation of the time pick-off of the anode pulse from the recovered pulse using EJ-309.}
  \label{fig:at_rt}
\end{figure}

Constant fraction discrimination (CFD) \cite{Fallu-Labruyere2007} was performed on the anode pulse and the recovered pulse respectively. The CFD time pick-off used an attenuation fraction of 1 with a 7.2 ns delay. Fig. \ref{fig:at_rt1} shows the scatter plot of the time-pickoff between the anode pulse and the recovered pulse. The same plot is also shown as the histogram of the difference between the anode timing and the recovered timing in Fig. \ref{fig:at_rt2}. The histogram has a bias with its mean at -67 $\pm$ 1.0 ps, and a standard deviation of 102 $\pm$ 1.5 ps; this standard deviation of 102 ps is the uncertainty in the estimate of the anode timing from the recovered pulse. Furthermore, we observed that the mean bias and standard deviation increased with decreasing charge, as Table \ref{table:1} illustrates.

\newcolumntype{M}[1]{>{\centering\arraybackslash}m{#1}}
\begin{table}[H]
\centering
\begin{tabular}{|M{0.8in}|M{0.5in}|M{0.9in}|} 
 \hline
 \thead{Energy range\\ (keVee)} & \thead{Mean\\ (ps)} & \thead{Standard\\ deviation (ps)} \\[0.5ex] 
 \hline
 80-150 & -90.62 & 155.57 \\ 
 150-200 & -83.94 & 119.85 \\
 200-300 & -68.09 & 95.70 \\
 300-400 & -58.16 & 76.63 \\
 400-500 & -52.88 & 66.90 \\
 500-600 & -49.79 & 61.59 \\ [1ex] 
 \hline
\end{tabular}
\caption{The mean and standard deviation of the distribution of the difference between the anode and the recovered timing when the events in Fig. \ref{fig:at_rt1} were divided into several energy ranges.}
\label{table:1}
\end{table}
The reason for the negative bias is that the recovered pulses after deconvolution appear to arrive slightly before the original anode pulses, most likely due to the amplified noise in the recovered pulse triggering the digitizer. 
Compared to the CFD time pick-off applied to the leading edge of the damped sinusoid discussed in \cite{Mishra2018}, the CFD time pick-off applied to the deconvolved pulses reduced the uncertainty on the timing by about 80\%.

\begin{figure}[H]
  \centering
  \begin{subfigure}[t]{0.75\linewidth}
    \includegraphics[width=\linewidth]{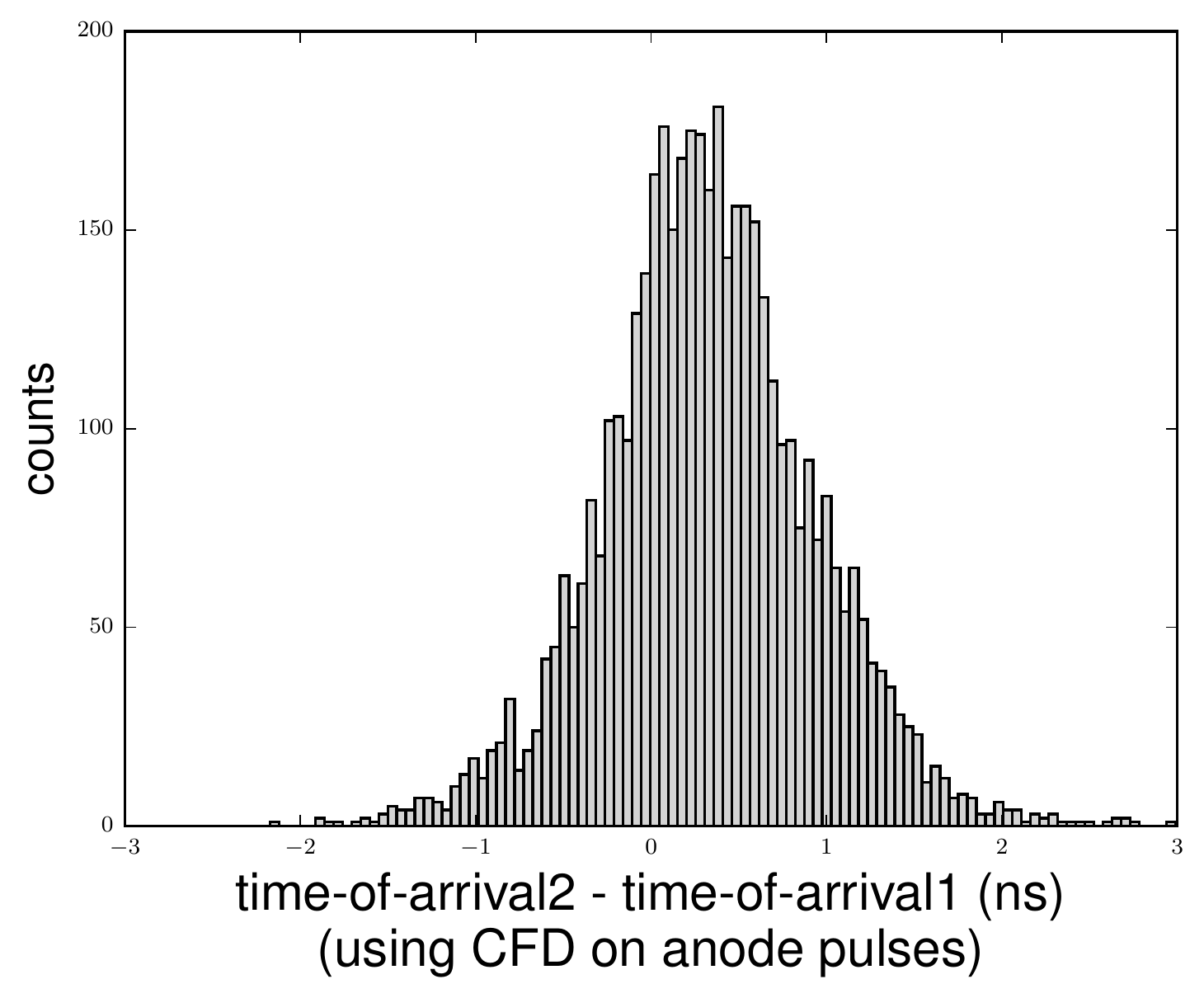}
     
    \caption{The histogram of the difference in the time of arrival of the coincident gamma rays using anode pulses ($\sigma = 603 \pm 18 ps$).}
    \label{fig:hist1}
  \end{subfigure}
  \hfill
  \begin{subfigure}[t]{0.75\linewidth}
    \includegraphics[width=\linewidth]{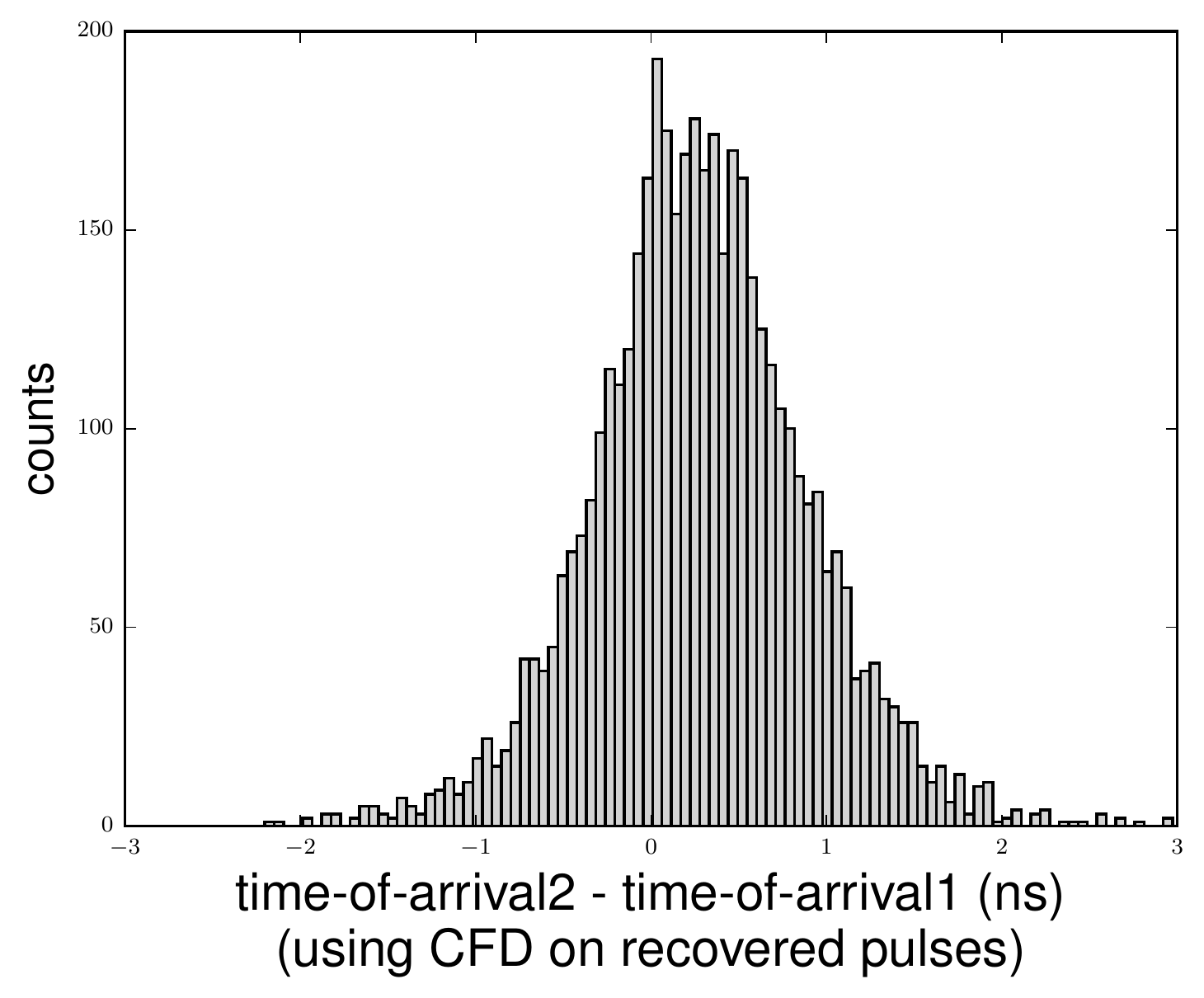}
    
    \caption{The histogram of the difference in the time of arrival of the coincident gamma rays using recovered pulses ($\sigma = 617 \pm 21 ps$).}
    \label{fig:hist2}
  \end{subfigure}
  \caption{Coincidence measurements with Na-22.}
  \label{fig:hist}
\end{figure}

We also performed coincidence measurements using two EJ-309 detectors to measure the uncertainty in the time pick-off of the anode pulses and the recovered pulses. We placed an Na-22 source between the two detectors, which were connected to the 7.00 MHz and 15.25 MHz resonators respectively. Each resonator was connected to a separate fan-in circuit to digitize sinusoids from coincident events. The anode pulses from the detectors and the corresponding damped sinusoids from the fan-ins were digitized simultaneously. We then calculated the time-of-arrival of the coincident pulses using CFD. Fig. \ref{fig:hist} shows the histogram of the difference in the time-of-arrival of the coincident events using the anode pulse ($\sigma$ = 603 $\pm$ 18 ps) and using the recovered pulse ($\sigma$ = 617 $\pm$ 21 ps). The increased timing uncertainty in Fig. \ref{fig:hist2} is the result of the additional uncertainty (see Fig. \ref{fig:at_rt}) introduced when the recovered pulses are used to calculate the time-of-arrival.

\subsection{Pulse-shape discrimination (PSD) using the recovered pulse} 
We used a Cf-252 source to induce pulses in an EJ-309 detector connected to the 7.00 MHz resonator in series with the fan-in circuit. The charge-integration method was used to discriminate the neutron pulses from the gamma pulses. 

\begin{figure}[H]
  \centering
  \begin{subfigure}[t]{0.75\linewidth}
    \includegraphics[width=\linewidth]{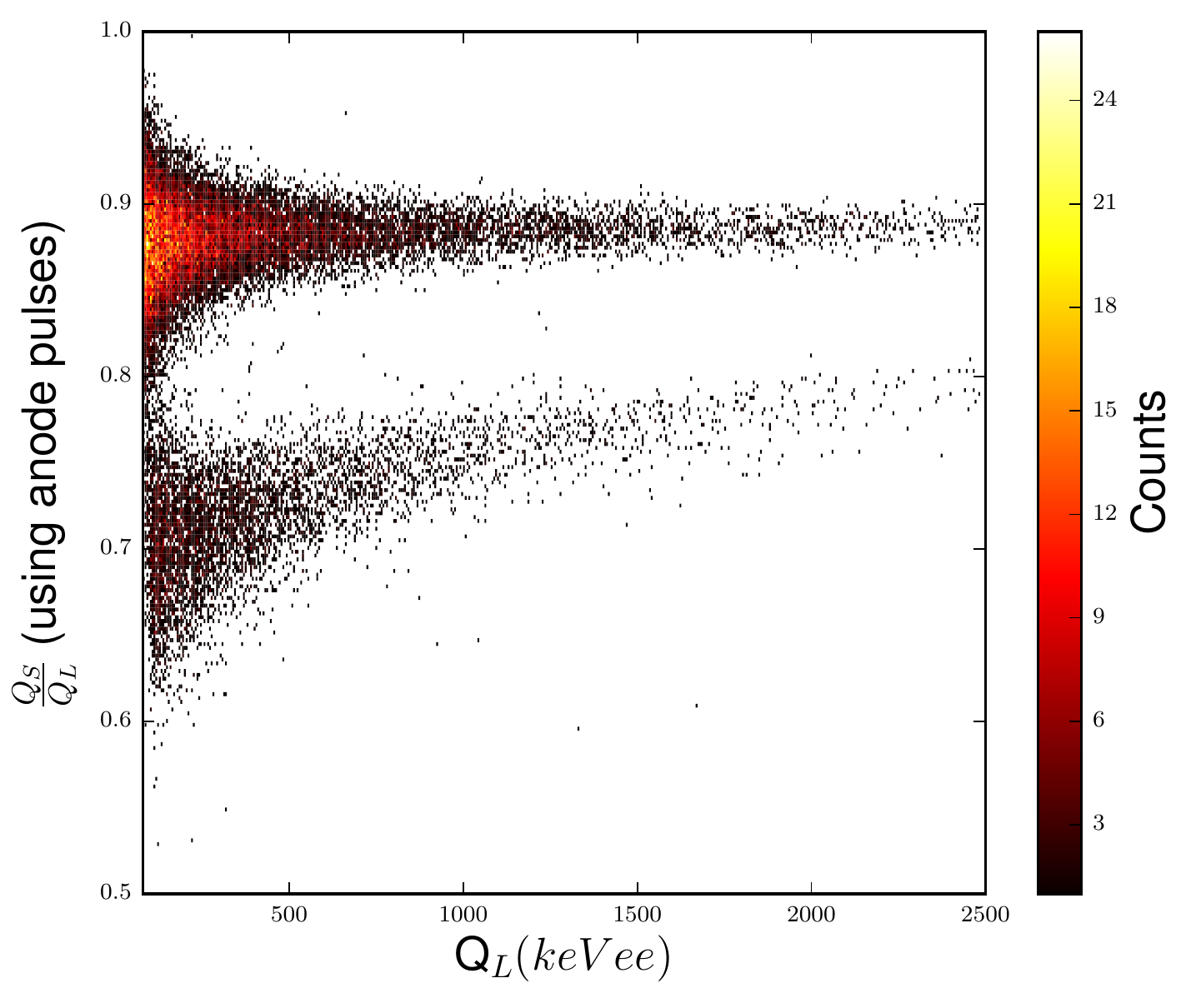}
     
    \caption{Distribution of the discrimination parameter against the total charge using the anode pulses.}
    \label{fig:psd1}
  \end{subfigure}
  \hfill
  \begin{subfigure}[t]{0.75\linewidth}
    \includegraphics[width=\linewidth]{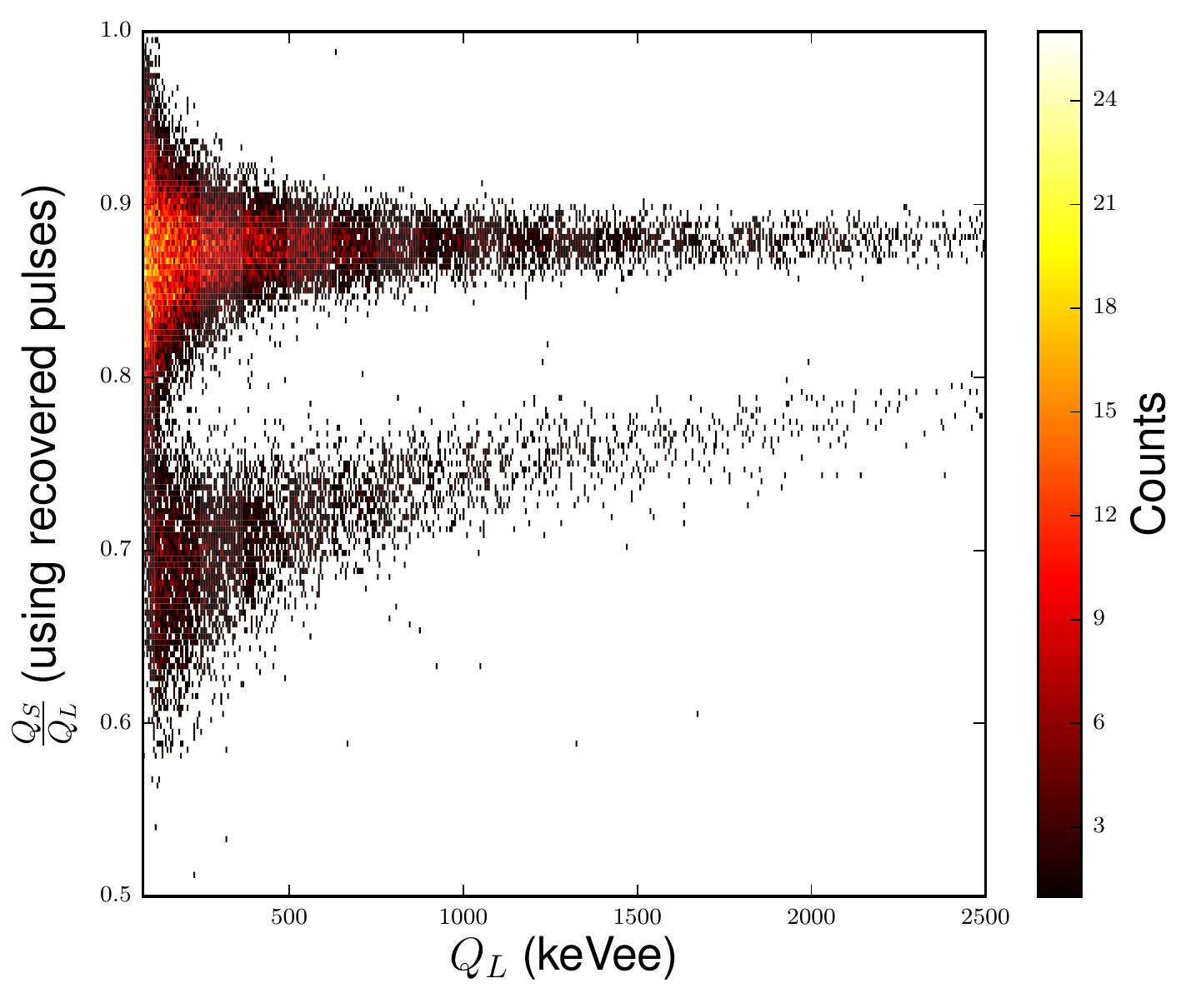}
    
    \caption{Distribution of the discrimination parameter against the total charge using the recovered pulses.}
    \label{fig:psd2}
  \end{subfigure}
  \caption{Pulse shape discrimination (color online).}
  \label{fig:psd}
\end{figure}

\begin{figure}[H]
  \centering
  \begin{subfigure}[t]{0.75\linewidth}
    \includegraphics[width=\linewidth]{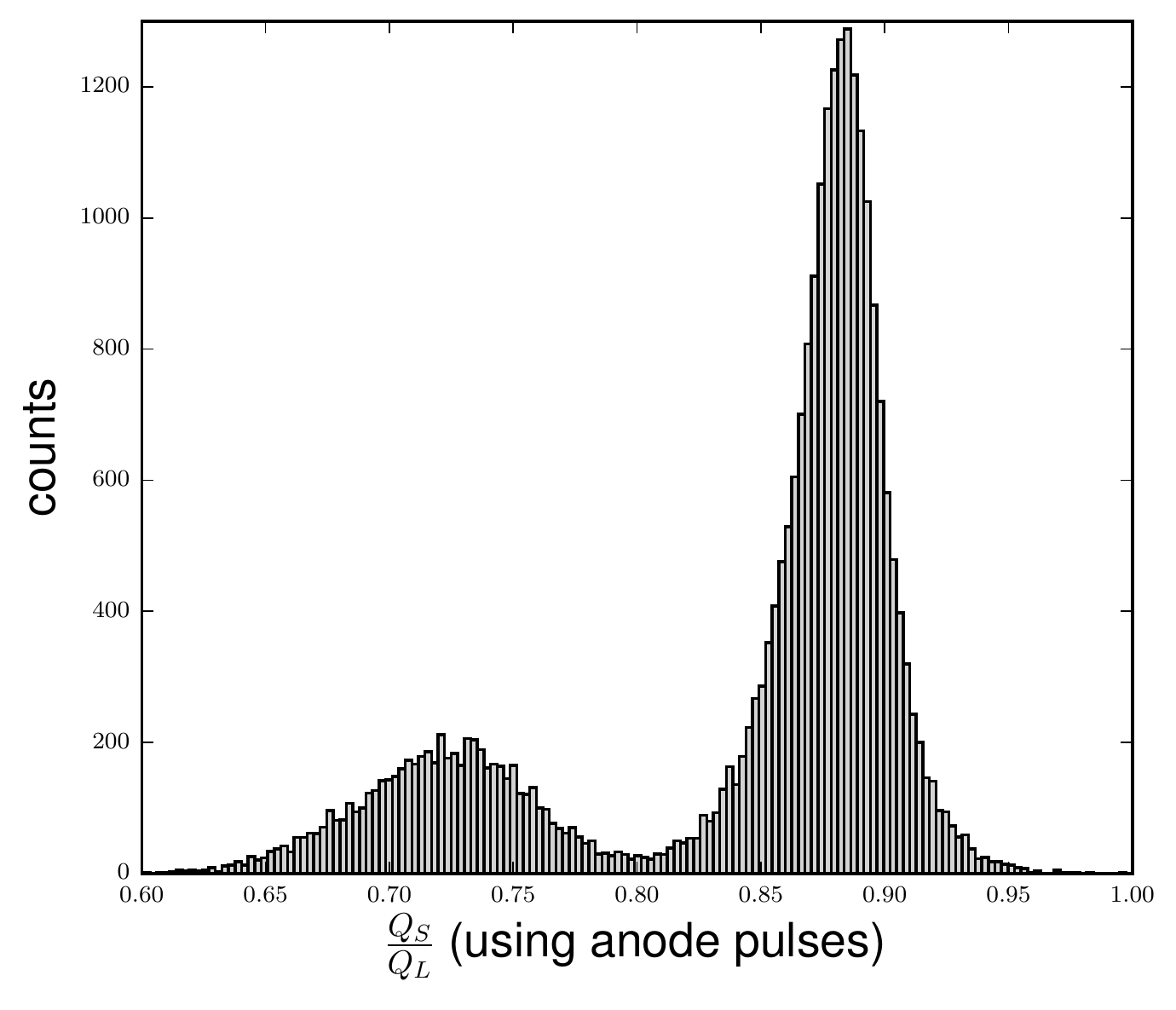}
     
    \caption{Distribution of the discrimination parameter when anode pulses were used (FOM = 1.2).}
    \label{fig:psdd1}
  \end{subfigure}
  \hfill
  \begin{subfigure}[t]{0.75\linewidth}
    \includegraphics[width=\linewidth]{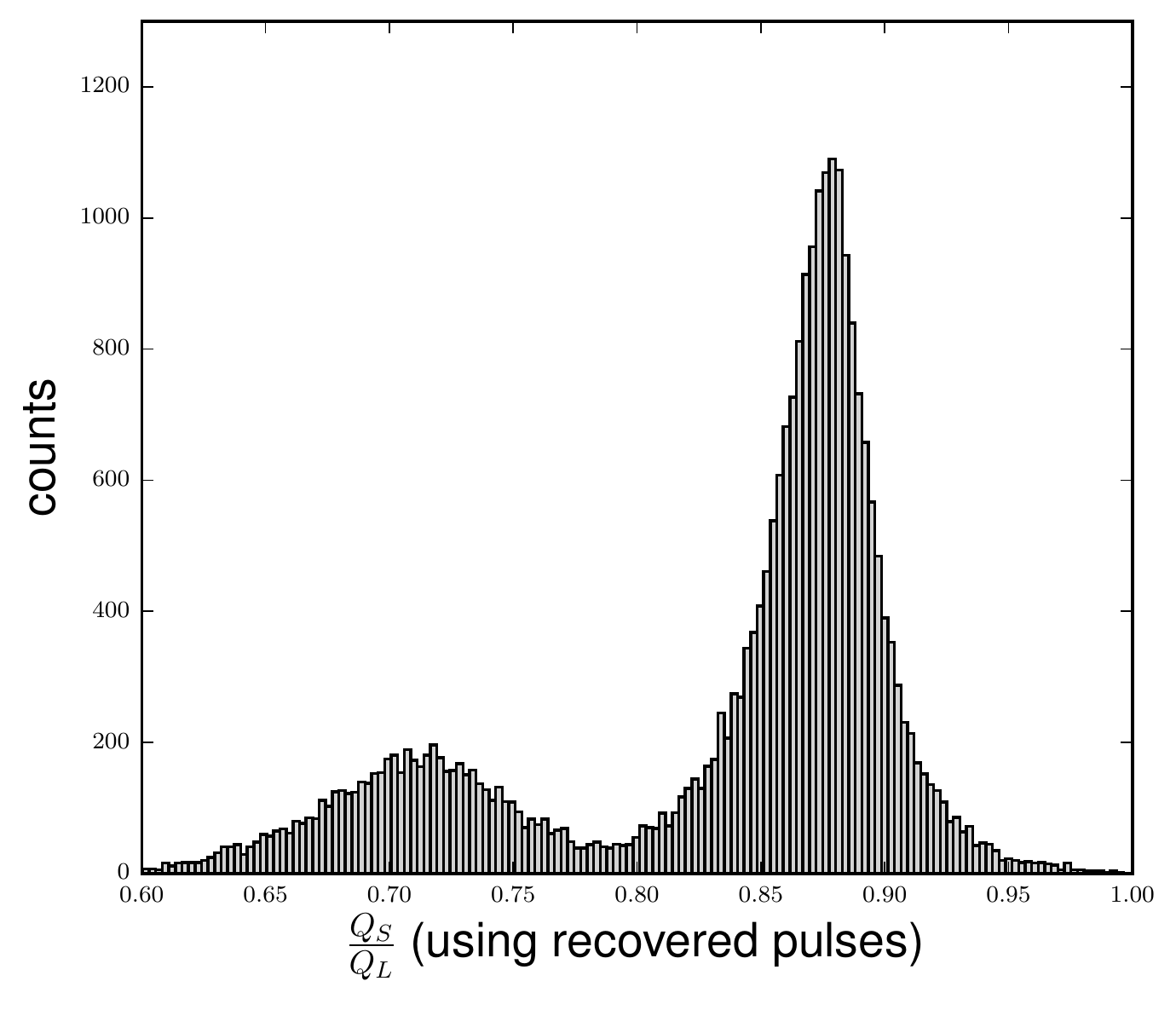}
    
    \caption{Distribution of the discrimination parameter when recovered pulses were used (FOM = 1.08).}
    \label{fig:psdd2}
  \end{subfigure}
  \caption{Comparison of the figure of merit.}
  \label{fig:psdd}
\end{figure}

The discrimination parameter was chosen as the ratio of the area under the anode pulse excluding its tail ($Q_{S}$), to the total area ($Q_{L}$) under the pulse. The start time of the pulse was fixed at six samples to the left of the pulse peak. The pulse length was kept constant at 120 samples. The stop time of the integration length for $Q_{S}$ was optimized by varying it from 0 to 20 samples and selecting the integration length that maximized the figure of merit (FOM) \cite{Winyard1971}. The figure of merit (FOM) was calculated using the ratio of the difference between the means of the two peaks to the sum of the full-width at half-maxima (FWHM) of the two peaks. 
\begin{equation}
  FOM = \frac{\mu_{gamma} - \mu_{neutron}}{FWHM_{gamma} + FWHM_{neutron}} 
\end{equation}
where $\mu$ denotes centroid and $FWHM$ denotes full-width at half-maximum. Fig. \ref{fig:psd} shows the 2D-histogram of the discrimination parameter $\frac{Q_{S}}{Q_{L}}$ against the total charge (area) $Q_{L}$ of the pulse for the anode and the recovered pulses respectively. The gamma band at the top and the neutron band at the bottom for the recovered pulses (Fig. \ref{fig:psd2}) show equally good separation compared to the anode pulses (Fig. \ref{fig:psd1}) for a threshold of 80 keVee. Fig. \ref{fig:psdd} shows the distribution of the discrimination parameter when the 2D-histograms shown in Fig. \ref{fig:psd} were summed onto the Y-axis. The peak with lower discrimination parameter corresponds to the neutrons and the peak with the higher parameter corresponds to the gamma-rays. 

We obtained a FOM of 1.20 using the anode pulses; it decreased slightly to 1.08 when recovered pulses were used most likely due to amplified noise on the baseline of the recovered pulses. The amplified noise at high frequencies makes it more difficult to precisely calculate the area under the tail of the recovered pulse to determine $Q_{L}$. This reduces the sensitivity of the discrimination parameter to the shape of the pulse, which reduces the FOM.   


\begin{figure}[H]
  \centering
  \begin{subfigure}[!htb]{0.65\linewidth}
    \includegraphics[width=\linewidth]{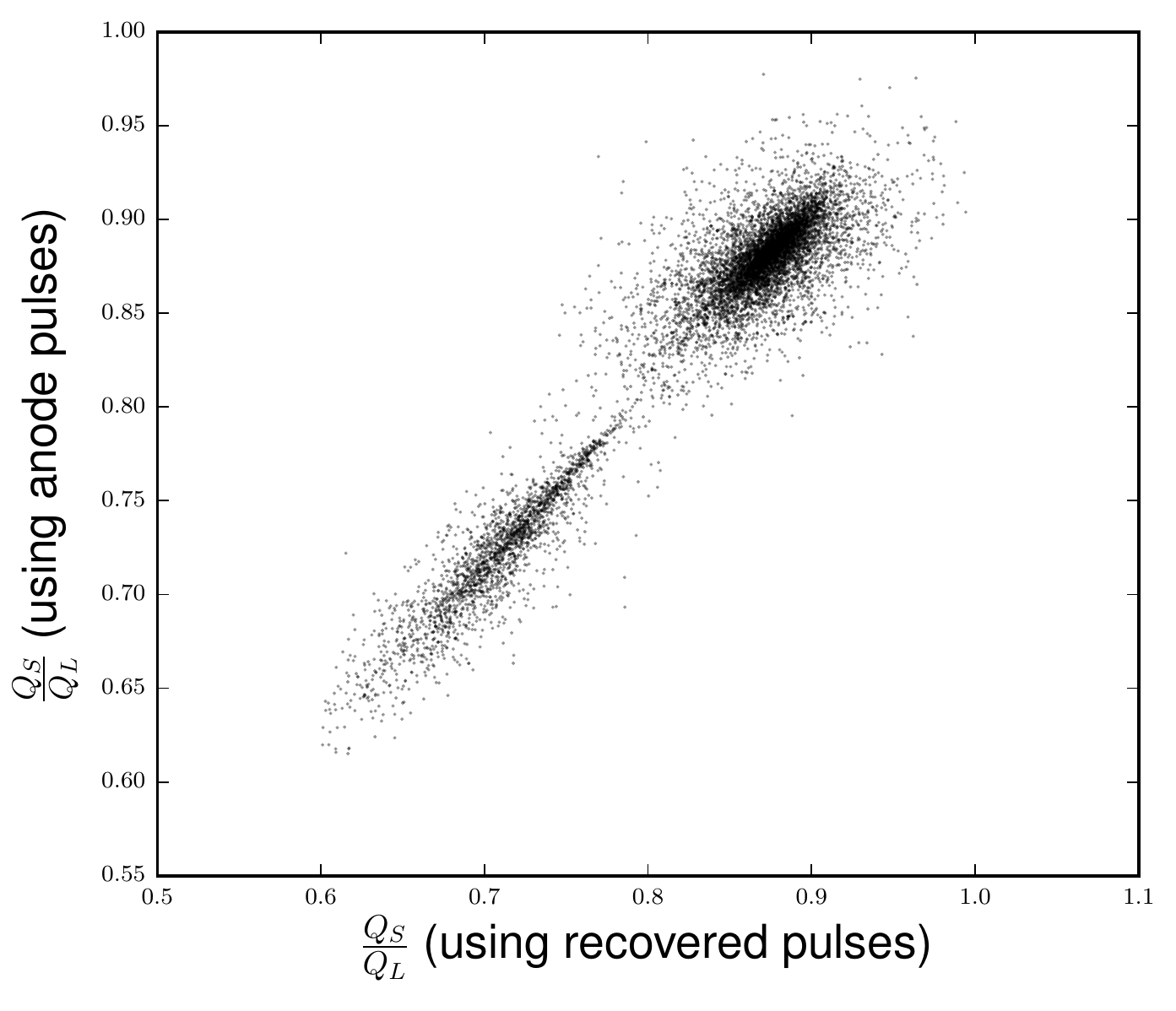}
     
    \caption{Scatter plot between the discrimination parameter from the anode pulse plotted against the discrimination parameter from the corresponding recovered pulse.}
    \label{fig:psdpsd}
  \end{subfigure}
  \hfill
  \begin{subfigure}[!htb]{0.65\linewidth}
    \includegraphics[width=\linewidth]{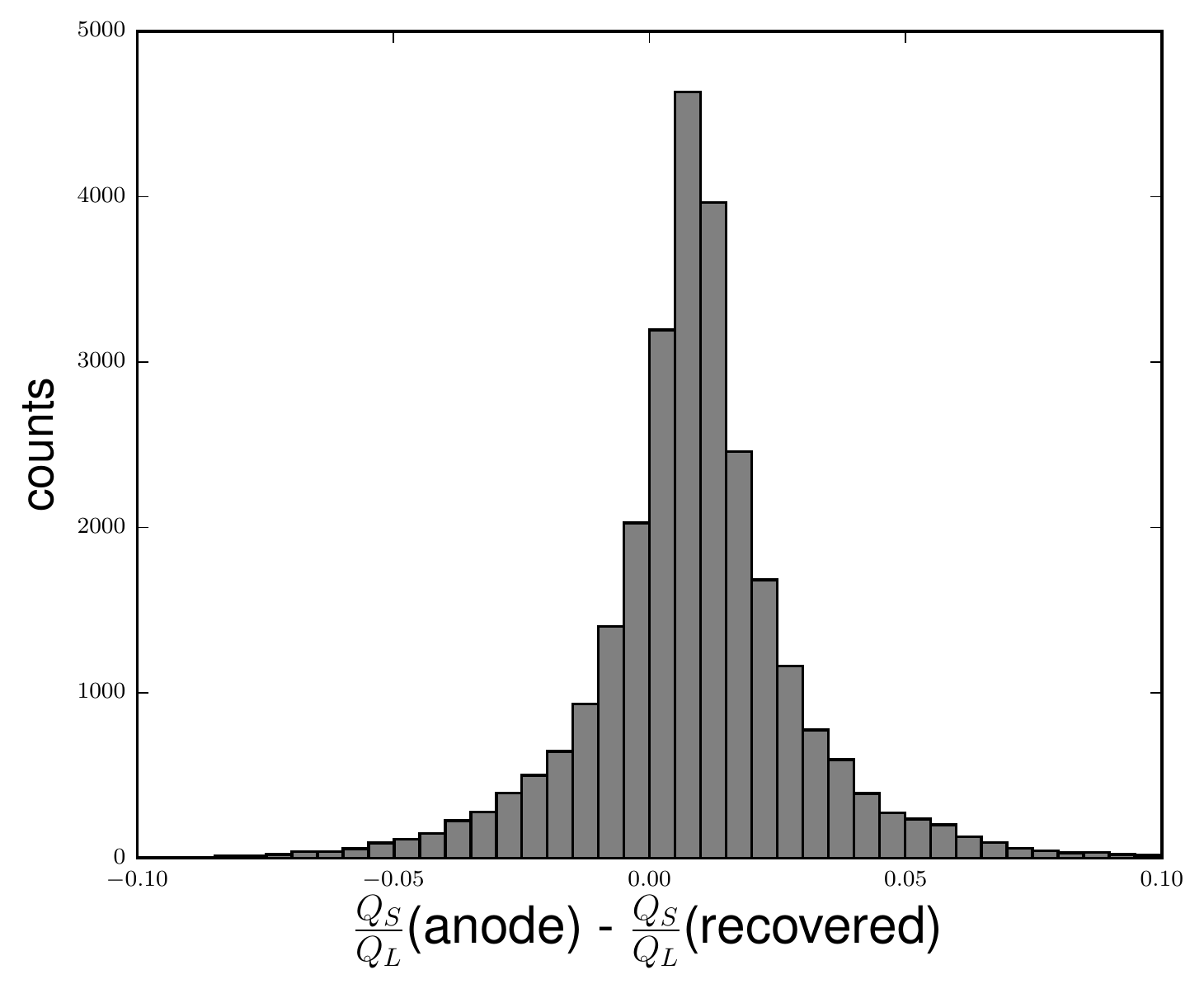}
    
    \caption{The events of the scatter plot in Fig. \ref{fig:psdpsd} shown as the histogram of the difference between the discrimination parameters using the anode and the recovered pulse with a standard deviation of 0.02 $\pm$ 0.00034.}
    \label{fig:psdpsd_hist}
  \end{subfigure}
  \caption{Comparison between the discrimination parameters using the anode pulse and the recovered pulse respectively.}
  \label{fig:psdpsdd}
\end{figure}

Fig. \ref{fig:psdpsd} shows the scatter plot between the discrimination parameters using the anode pulse and the recovered pulse respectively. Fig. \ref{fig:psdpsd_hist} shows the events of the scatter plot as the histogram of the difference between the discrimination parameters using the anode and the corresponding recovered pulse. The standard deviation in the distribution was computed to be 0.02 $\pm$ 0.00034.

\section{Conclusions and future work}
We demonstrated frequency domain multiplexing of two EJ-309 organic scintillator detectors by convolution and deconvolution. The charge collected under the anode pulse can be estimated from the pulse recovered by deconvolution with an uncertainty of about 4.4 keVee. The time of arrival of the anode pulse can be estimated from the recovered pulse with an uncertainty of about 102 ps in addition to the inherent timing uncertainty associated with the original anode pulse. A CeBr\textsubscript{3} inorganic scintillator was also connected to the multiplexer to measure the Cs-137 gamma spectrum; a standard deviation of 13.8 keV was observed at 662 keV using the deconvolved pulses compared to a standard deviation of 13.5 keV when the original pulses were used. Coincidence measurements with Na-22 resulted in a timing uncertainty of 603 ps using the original detector pulses that increased to 617 ps using the recovered pulses.  

Pulse shape discrimination performed on the recovered pulses showed a small decrease in the FOM. A FOM of 1.08 was observed when the charge integration method was applied on the recovered pulses compared to 1.2 when the original pulses were used. The uncertainty in the distribution of the difference between the discrimination parameters calculated from the anode pulse and the corresponding recovered pulse was found to be 0.02, so the change in FOM is significant. It is most likely caused by amplified noise in the baseline of the recovered pulses. 

We are currently working on the reconstruction of the original anode pulses when more than one detector fires in the same digitizer record. In future work, we will focus on the problem of signal recovery when the deconvolution is performed on a damped sinusoid of one particular frequency in the presence of other sinusoids of different oscillation frequencies.   

\section{Acknowledgements}
This work was sponsored in part by the NNSA Office of Defense Nuclear Nonproliferation R\&D through the Consortium for Verification Technology (CVT), grant number DE-NA0002534.

\section*{References}

\bibliography{library}

\end{document}